**Approaching the robust linearity in dual-floating van der Waals photodiode**

*Jinpeng Xu, Xiaoguang Luo*, Xi Lin, Xi Zhang, Fan Liu, Yuting Yan, Siqi Hu, Mingwen Zhang, Nannan Han*, Xuetao Gan*, Yingchun Cheng*, Wei Huang**

Dr. J. Xu, Dr. X. Luo, X. Lin, X. Zhang, F. Liu, Y. Yan, Dr. N. Han, Prof. W. Huang
Frontiers Science Center for Flexible Electronics (FSCFE), Shaanxi Institute of Flexible Electronics (SIFE) & Shaanxi Institute of Biomedical Materials and Engineering (SIBME), Northwestern Polytechnical University, Xi'an 710129, China
*E-mail: iamxgluo@nwpu.edu.cn, iamnhan@nwpu.wdu.cn, iamwhuang@nwpu.edu.cn

Dr. S. Hu, M. Zhang, Prof. X. Gan
Key Laboratory of Light Field Manipulation and Information Acquisition, Ministry of Industry and Information Technology, and Shaanxi Key Laboratory of Optical Information Technology, School of Physical Science and Technology, Northwestern Polytechnical University, Xi'an 710129, China
*E-mail: xuetaogan@nwpu.edu.cn

Prof. Y. Cheng
Key Laboratory of Flexible Electronics & Institute of Advanced Materials, Jiangsu National Synergetic Innovation Center for Advanced Materials, Nanjing Tech University, Nanjing 211816, China
*E-mail: iamyccheng@njtech.edu.cn

Dr. J. Xu
Institute of Physics, Henan Academy of Sciences, Zhengzhou 450046, PR China

**Abstract**

Two-dimensional (2D) material photodetectors have gained great attention as potential elements for optoelectronic applications. However, the linearity of the photoresponse is often compromised by the carrier interaction, even in 2D photodiodes. In this study, we present a new device concept of dual-floating van der Waals heterostructures (vdWHs) photodiode by employing ambipolar $MoTe_2$ and *n*-type $MoS_2$ 2D semiconductors. The presence of type II heterojunctions on both sides of channel layers effectively deplete carriers and restrict the photocarrier trapping within the channel layers. As a result, the device exhibits robust linear photoresponse under photovoltaic mode from the visible (405 nm) to near-infrared (1600 nm) band. With the built-in electric field of the vdWHs, we achieve a linear dynamic range of ~ 100 dB, responsivity of ~ 1.57 A/W, detectivity of ~ $4.28 \times 10^{11}$ Jones, and response speed of ~ 30 μs. Our results showcase a promising device concept with excellent linearity towards fast and low-loss detection, high-resolution imaging, and logic optoelectronics.

## 1. Introduction

Photodetectors, which convert light into electric signal, are the building blocks of modern communications, environmental monitoring, machine version, artificial intelligence, etc[1-3]. With the development of processing technology and commercial demand, high-performance photodetections are required in terms of sensitivity, speed, linear photoresponse, broadband detection, integration, and power consumption[4-6]. The core elements of photodetectors are the photoactive materials and structures. Nowadays, a plethora of inorganic and organic semiconductors have been utilized to explore their photodetection ability. Some new materials or structures, such as two-dimensional (2D) semiconductors, perovskites, quantum dots, and nanowires, are considered as the candidates for next-generation photodetectors[6-8]. Among them, layered 2D semiconductors are widely used in photodiodes and phototransistors due to their high carrier mobility, favorable bandgap, flexible property, and large light absorption cross-section[9-10]. In particular, the atomically thin nature with hanging-bond-free surfaces allows 2D semiconductors to be constructed and integrated by the architecture of van der Waals heterostructures (vdWHs)[11], which promote photocarrier separation and extend response band during photodetection[12-13].

Linear photoresponse with respect to light intensity is a vital indicator for reliable and high-resolution identification of optical signals. Generally, achieving robust linearity through 2D phototransistors is challenging because of the inevitable carrier trapping in the atomically thin channel[14-16], while it seems an easy implementation for 2D photodiodes with the promoted photocarrier separation by built-in electric field[17-20]. In fact, an ideal photodiode performs the robust linear photoresponse owing to the minimal interaction of photocarriers within the junction. However, the doping level or the carrier regulation uniformity in an as-fabricated 2D photodiode are usually out of expectation. For instance, carrier distributions near the top and bottom interfaces of the photodiode channel are typically different, which probably results in the distinct nonlinear photoresponse[21]. Applying electric field on the 2D channel is an effective way to control the carrier behavior, such as the built-in electric field formed in multijunction heterostructure[22]. With the assistance of the built-in electric field, carriers are effectively depleted in the junction or transferred out of the channel, rather than trapped in the bandgap. As a result, the capture and release of carriers are greatly restrained during transport, facilitating the linear photoresponse. It has been reported that photodetection performance can be remarkably improved by an extra floating layer, however, with the nonlinear photoresponse[23]. The probable explanation for this nonlinearity is that the carriers in the junction are not completely depleted via a single floating layer. Therefore, the 2D multijunction heterostructure with double floating layers deserves the attention for designing high-performance photodiode with linear photoresponse.

In this study, we report a new device concept of dual-floating vdWHs photodiode based on ambipolar $MoTe_2$ (bandgap ~ 1.0 eV) and *n*-type $MoS_2$ (bandgap ~ 1.2 eV). The $MoTe_2/MoS_2$ vdWH channel of the photodiode was fabricated by dry transfer technique[24] and sandwiched between top $MoS_2$ and bottom $MoTe_2$ floating layers, as shown schematically in Figure 1a. The built-in electric field around dual-floating layers effectively suppressed carriers in each channel from the top and bottom sides, resulting in a robust linear photoresponse under the photovoltaic mode, even for the non-ideal photodiode. Attributed to the intralayer and interlayer excitations in the vdWHs, the broadband photodetection, ranging from visible to near-infrared (NIR) band (405-1600 nm), was achieved at room temperature. With 532 nm illumination, we obtained high-performance results (including a linear dynamic range of ~ 100 dB, responsivity of ~ 1.57 A/W, detectivity of ~ $4.28 \times 10^{11}$ Jones, and response speed of ~ 30 μs) with this dual-floating vdWHs photodiode. Our findings provide a promising route for designing high-performance 2D photodiode with excellent linear photoresponse.

## 2. Results and Discussion

### 2.1. Device fabrication and characterization

2D ambipolar $MoTe_2$ and *n*-type $MoS_2$ were mechanically exfoliated for the fabrication of dual-floating vdWHs photodiode that through dry transfer method[24]. Au films with thickness of ~ 50 nm were transferred on the two middle layers ($MoS_2$ and $MoTe_2$) as the source and drain electrodes, respectively, to weaken the Fermi level pinning effect at the metal/semiconductor interface via van der Waals contact[25]. Additional fabrication details can be found in the Experimental Section and Figure S1. Figure 1a illustrates the schematic of the dual-floating vdWHs photodiode on a $SiO_2$/Si substrate. The middle $MoTe_2/MoS_2$ vdWH acts as the conducting channel, sandwiched by $MoS_2$ (top floating) and $MoTe_2$ (bottom floating) layers. Thus, the top $MoS_2$ and bottom $MoTe_2$ can be regarded as the dual-floating layers in this vdWHs. The hexagonal boron nitride (hBN) nanosheet, which holds the whole $MoS_2/MoTe_2/MoS_2/MoTe_2$ dual-floating vdWHs, provides a clean substrate that prevents carrier trapping and surface bonding from the $SiO_2$ surface. In this study, we conducted electrical and optoelectronic characterizations on three $MoS_2/MoTe_2/MoS_2/MoTe_2$ dual-floating vdWHs photodiodes (denoted as Device A, B, and C), with the size parameters listed in Table S1.

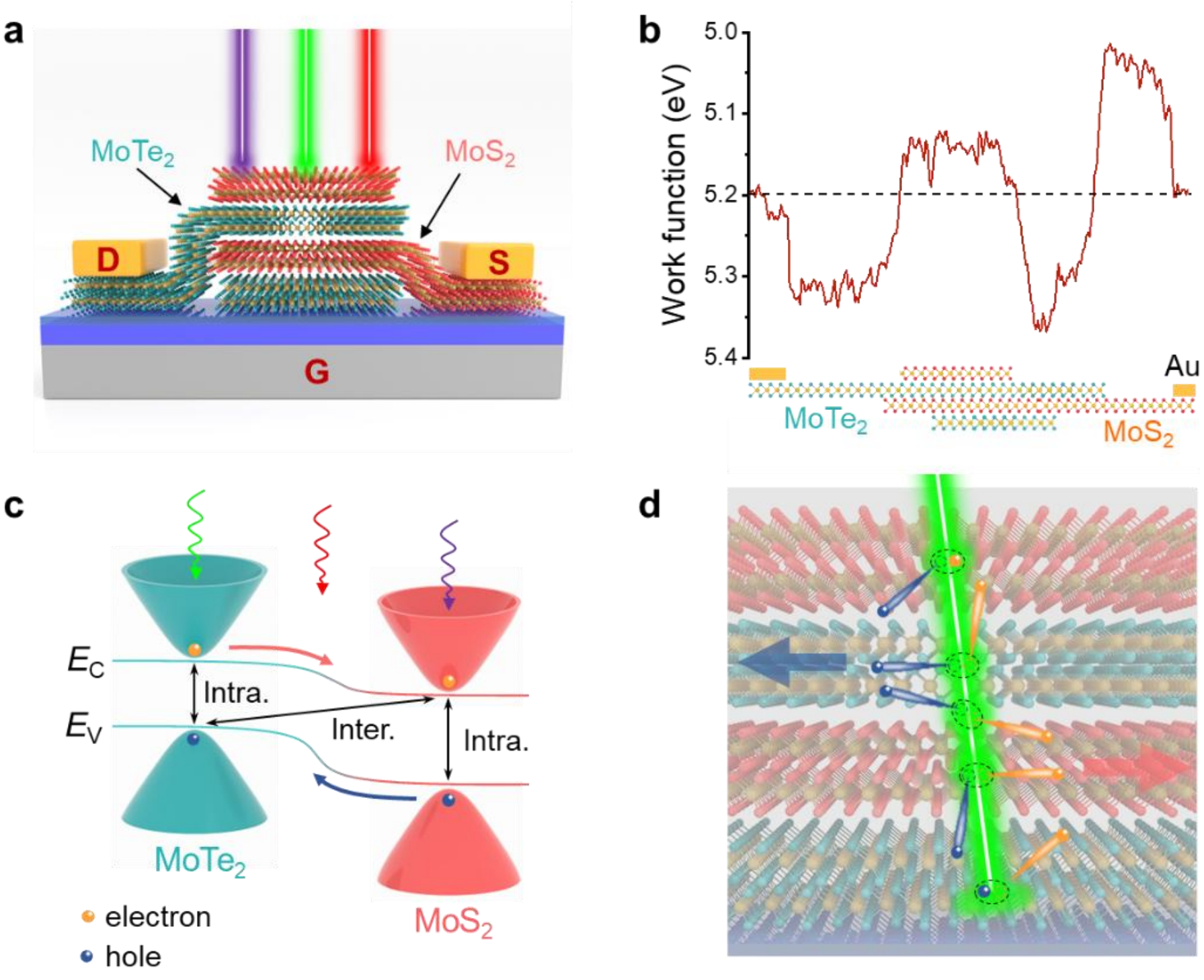

**Figure 1. The $MoS_2/MoTe_2/MoS_2/MoTe_2$ dual-floating vdWHs photodiode. a**) Schematic of the dual-floating vdWHs photodiode, where the middle $MoTe_2/MoS_2$ heterostructure (conducting channel) is sandwiched by $MoS_2$ (top) and $MoTe_2$ (bottom) layers on the hBN/$SiO_2$/Si substrate. S: source, D: drain, G: gate. All 2D materials are multilayer in actual devices. **b**) Work functions along the surface of Device A (with thicknesses of 3.5/3.6/6.5/48 nm for $MoS_2/MoTe_2/MoS_2/MoTe_2$) from the KPFM results of Figure S2d. **c**) Band structure diagram of the $MoTe_2/MoS_2$ heterostructure and three kind excitations (two intralayer and one interlayer) therein. $E_F$: Fermi level, $E_C$: conductance band, $E_{V.}$: valence level. **d**) Diagram of carrier dynamics in the dual-floating vdWHs photodiode.

The thickness of each layer and the surface work function of the dual-floating vdWHs were identified using atomic force microscopy (AFM) and its Kelvin probe force microscopy (KPFM) mode. Figures 1b and S2 present the results of a focused device (Device A) in this work. It was confirmed that all 2D materials in the photodiode were multilayer, indicating the similar band structures for two $MoTe_2$ (or $MoS_2$) layers. The work functions of $MoTe_2$ and $MoS_2$ were determined to be approximately 5.32 eV and 5.03 eV, respectively, by choosing the work function of Au (~ 5.2 eV) as a reference. The surface work function of the vdWHs was found to be larger than that of the homogeneous $MoS_2$ layer, confirming the effective electron transfer from $MoS_2$ to $MoTe_2$ upon contact. To assess the quality of the heterojunction channel, Raman and photoluminescence (PL) spectra were examined with 532 nm laser excitation, as shown in Figure S3. The intrinsic Raman peaks of both $MoS_2$ and $MoTe_2$ (the in-plane vibrational mode $E^1_{2g}$ and the out-of-plane vibrational mode $A_{1g}$) appear distinctly in the Raman spectrum of the vdWHs, suggesting the high quality of the heterostructure[26-27]. The PL spectra of the homogeneous $MoS_2$ layer and the vdWHs are shown in Figure S3b, where two identified peaks (indexed by $A_1$ and $B_1$) are assigned to the direct excitonic transitions of $MoS_2$ due to the energy split of valence band by spin-orbital coupling[28]. Both Raman and PL spectra indicate high-quality interfaces, as well as very few impurities that brought in during the device fabrication. In-situ PL spectra were further analyzed under different gate voltages. The PL spectrum of the homogeneous $MoS_2$ layer (with $A_1$ and $B_1$ peaks at approximately 1.817 eV and 2.0 eV, respectively) was barely affected by gating effect, suggesting heavy $n$-type doping[29-30]. Nevertheless, the PL spectrum of the vdWHs was remarkably tunable with gate voltage $V_g$. A slight red-shift and a notable PL enhancement were observed when $V_g$ varies from 0 to -60 V. The former may be attributed to the electric-field-regulated bandgap[31], while the latter was explained by the phase-space filling effect[29-30, 32] where the exciton oscillator strength could be reduced with the increase of electron density due to the Pauli exclusion principle[30]. It is learnt in the literature that the gate-tunable PL effect of $MoS_2$ exclusively manifests at low electron density level[29-30], and therefore the electron density at the $MoS_2$ layer in the vdWHs is notably lower when compared to the homogeneous $MoS_2$ layer. Consequently, one can deduce again that electron transfer within the vdWHs predominantly occurs from $MoS_2$ to $MoTe_2$ upon contact.

**2.2. Electrical properties of the photodiode**

The conduction band minimum and the valence band maximum of multilayer $MoTe_2$ ($MoS_2$) are approximately –3.8 eV (–4.2 eV) and –4.8 eV (–5.4 eV), respectively[21, 33]. Combined with the KPFM results, i.e., the work functions of $MoTe_2$ ~ 5.32 eV and $MoS_2$ ~ 5.03 eV, the type II band alignment of $MoTe_2$/$MoS_2$ heterostructure can be inferred based on the charge transfer behavior, as shown Figure 1c. The band bending occurs at the interface and forms a depletion region[34]. Our dual-floating vdWHs have three type II heterojunctions at three $MoS_2$/$MoTe_2$ interfaces. The interlayer excitation at the $MoTe_2$/$MoS_2$ junctions can be exploited for NIR photodetection beyond the intralayer excitations in both $MoTe_2$ and $MoS_2$ layers[12, 35]. Under illumination, the photogenerated electron-hole pairs are separated effectively by the built-in electric field in type II heterojunctions, resulting in the electron accumulation in $MoS_2$ layers and hole accumulation in $MoTe_2$ layers (Figure 1d). The middle $MoTe_2$/$MoS_2$ vdWH channel is particularly conducive to charge accumulations because of the double-sided action. Figuratively, an “electron valley” and a “hole valley” are formed in $MoS_2$ and $MoTe_2$ conducting channels, respectively.

The electrical properties of the device were characterized in a vacuum chamber at room temperature, unless otherwise specified. The drain-source bias voltage ($V_{\text{ds}}$) was applied to the $MoTe_2$ channel through the Au electrode, the $MoS_2$ channel was grounded, and the gate voltage ($V_{\text{g}}$) was applied by the $SiO_2$/Si substrate. Figure S4a shows the transfer curves (i.e., source-drain current $I_{\text{ds}}$ versus $V_{\text{g}}$ at the given $V_{\text{ds}}$) of Device A, exhibiting the typical anti-ambipolarity of $MoTe_2$/$MoS_2$ heterostructure diode[36]. Four identifiable regimes with respect to gate voltage confirms the effective regulation of Fermi levels for both ambipolar $MoTe_2$ and $n$-type $MoS_2$ layers[34, 37]. The output curves (i.e., $I_{\text{ds}}$ versus $V_{\text{ds}}$ at the given $V_{\text{g}}$, see Figure S4b) reveal the gate-tunable rectification effect with the rectification ratio ranging from $10^0$ to $10^3$. Such low rectification ratio is evidence of the non-ideal photodiode, which can also be concluded by the fitted ideal factor ($n$) from the formula of

$$I_{\text{ds}} = I_{\text{S}}\left[\exp\left(\frac{eV_{\text{ds}}}{nkT}\right) - 1\right], \quad (1)$$

where $I_{\text{S}}$ is the reverse saturation current, $e$ is the elementary charge and $kT$ is the thermal energy. For $V_{\text{ds}} > 0$ at different gate voltages (Figure S4c), all ideal factors are greater than unity, implying that charge recombination or diffusion is available for both majority carriers and minority carriers in $MoTe_2$/$MoS_2$ channel. For $V_{\text{ds}} < 0$, three different configurations are found at different gate voltages (Figures S4d-S4f), including forward-rectifying diode, backward-rectifying diode, and Zener diode. Similar reconfigurations have also been observed in BP/$MoS_2$ vdWH diodes[38]. The gate-tunable band alignment is likely responsible for the diode reconfiguration, which determines the validity of band-to-band tunneling, as demonstrated in Figures S4g-S4i. In any case, our device is a non-ideal diode with type II heterojunctions.

### 2.3. Optoelectronic characteristics and mechanism analysis

The photodetection performance of Device A was initially evaluated under global illumination with a 532 nm laser spot (5 mm in diameter), in order to eliminate the photothermoelectric effect induced by temperature gradient from non-uniform illumination[39]. The transfer curves (Figure S5a) and output curves (Figure S5b) were measured again under illumination at different laser power densities ($P_{\text{in}}$). For reverse bias at $V_{\text{ds}} =- 0.1$ V, $I_{\text{ds}}$ increases with the laser power density ($P_{\text{in}}$) in the $V_{\text{g}}$ range from –80 to 0 V, indicating a clear photoresponse of the device. Focusing on the output curves at $P_{\text{in}} = 87.5$ mW/cm$^2$, the presence of short-circuit current ($I_{\text{sc}}$) and open-circuit voltage ($V_{\text{oc}}$) at different $V_{\text{g}}$ points out the main contribution of photovoltaic effect. The maximum $V_{\text{oc}}$ of our devices was achieved in the *p-n* configuration, specifically, around $V_{\text{g}} =- 50$ V for Device A, as depicted in Figure S5b. It is noteworthy that all of our devices exhibited a *p-n* configuration at $V_g = -50$ V, which is consequently maintained for all subsequent measurements, unless explicitly stated otherwise. Figure 2a displays the output curves with increasing $P_{in}$, and the increase in $I_{ds}$ at reverse bias is consistent with the results from transfer characteristics measurements. The electrical production capability was evaluated using the electrical power $P = I_{ds}V_{ds}$, from which the maximum power of our photodiode was verified in the level of nanowatts (Figure 2b). The power conversion efficiency ($PCE$) was calculated using the maximum electrical power, i.e., $PCE = P_{max}/P_{in}A$, where $A \sim 123.5$ μm$^2$ was the photoactive area of the vdWHs. As $P_{in}$ increased from 0.0008 to 87.5 mW/cm$^2$, $PCE$ initially increased to 11.1%, and then decreased to 8.1%, which is evidently higher than the record-high value of 9% for $MoS_2$/AsP vdWH photodiode[40]. Such a high $PCE$ and its tendency with respect to $P_{in}$ can be explained by the contribution of three type II heterojunctions in the vdWHs. With the increase of $P_{in}$, the laser penetrates deeper, and the contribution to photoresponse from bottom junctions gradually rises before reaching satuaration. After that, however, $PCE$ decreases with the further heightened illumination due to the limited capacity of the whole structure at light harvesting and energy conversion.

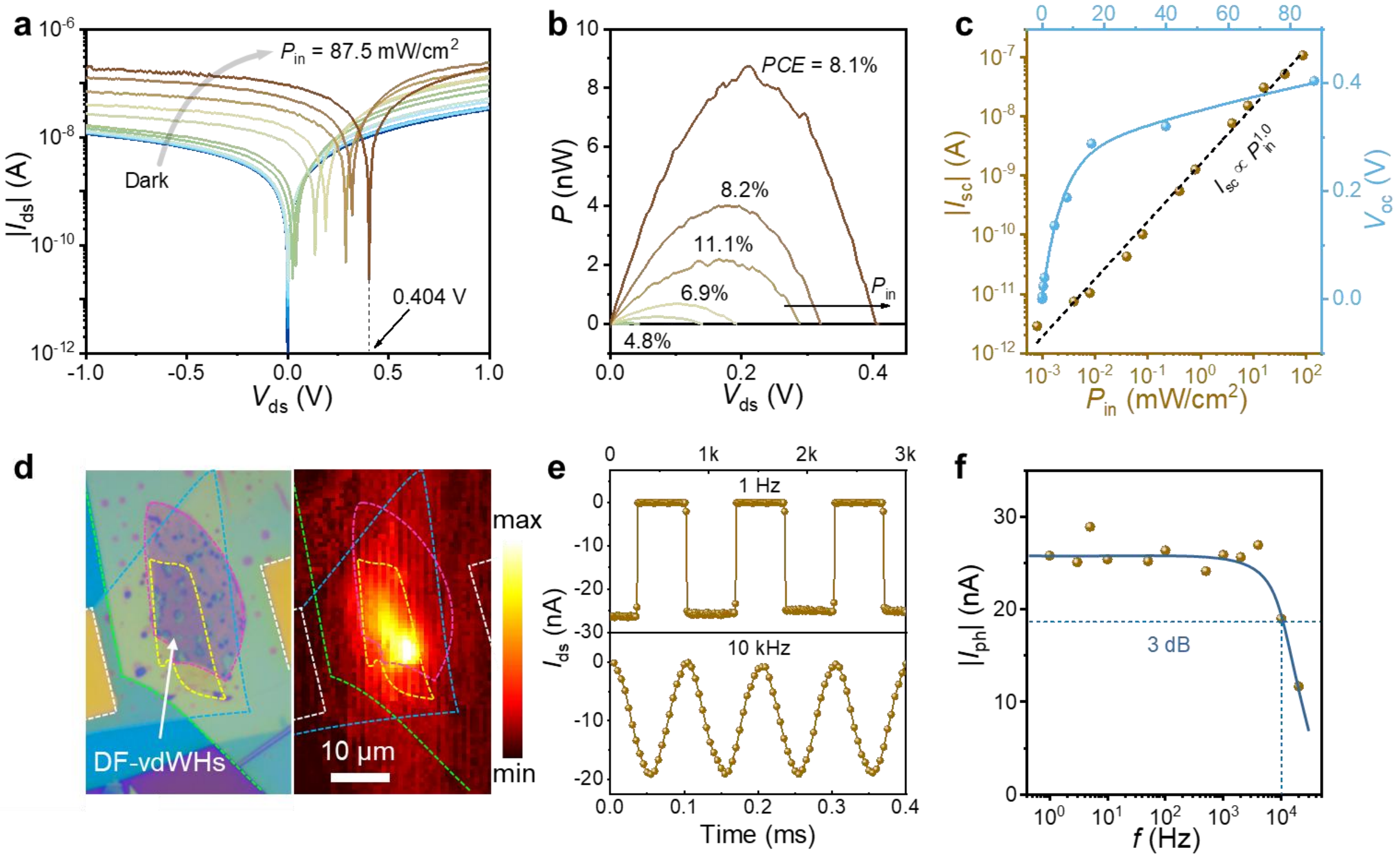

**Figure 2. Photodetection performance of the dual-floating vdWHs photodiode.** **a**) Output characteristic curves of Device A under different laser power densities. **b**), **c**) Extracted electrical power, $PCE$, $I_{sc}$ and $V_{oc}$ from output curves, the $LDR$ is ~100 dB. **d**) Optical micrograph of Device B (left panel, with thicknesses of 4/11/11/28 nm for $MoS_2/MoTe_2/MoS_2/MoTe_2$) and the scanning photocurrent image (right panel) when $V_{ds} = 0$ V, where the laser power is about 10 nW and the laser spot size is about 3 μm, the pink, green, blue, yellow, and white dashed lines outline the bottom $MoTe_2$, middle $MoS_2$, middle $MoTe_2$, top $MoS_2$, and Au electrodes, respectively, and the dual-floating vdWHs is labelled as DF-vdWHs. **e)** Transient photocurrent of Device B at the laser modulation frequency of 1 Hz (upper panel) and 10 kHz (lower panel) when $V_{ds} = 0$ V and $P_{in} = 42.4$ mW/cm$^2$. **f)** 3 dB bandwidth of the photodiode based on a series of measurements of transient photocurrent. The laser wavelength is 532 nm and the gate voltage is fixed at $V_g = -50$ V.

The photovoltaic effect was assessed through measurements of $I_{sc}$ and $V_{oc}$, as shown in Figure 2c. $V_{oc}$ follows a logarithmic dependence on $P_{in}$ with a maximum value of 0.404 V at $P_{in} = 87.5$ mW/cm$^2$, while $I_{sc}$ follows an excellent linear dependence with a maximum value of 110 nA. The photocurrent, defined as the difference between the current measured under illumination and in the dark (i.e., $I_{ph} = I_{illum.} - I_{dark}$), also follows a linear dependence on laser power density. In the absence of drain-source bias, the dark current ($I_{dark}$~0.1 pA) for our photodiode is quite low and the current under illumination $I_{illum.}$~$I_{sc}$. The linear region can be quantified by the linear dynamic range ($LDR$) in dB, given by:

$$LDR = 20 \log\left(\frac{P_h}{P_l}\right), \quad (2)$$

where $P_{h/l}$ is the highest/lowest laser power density of the linear region. Based on the data in Figure 2c, $LDR = 100$ dB is achieved for Device A. It is worth noting that 100 dB is still underestimated due to the limited illumination conditions. The reproducibility of the performance was found to be excellent from device to device. For another dual-floating vdWHs photodiode (Device B), the maximum $V_{oc}$ and $I_{sc}$ at $P_{in} = 45.6$ mW/cm$^2$ are 0.37 V and 97.8 nA, respectively, and the photoresponse also exhibits linear dependence on $P_{in}$, as shown in Figure S6. Based on the photocurrent mapping presented in Figure 2d, it can be inferred that the dominating light harvesting and photocarrier separation occur at the overlapped portion of the dual-floating vdWHs, rather than at the semiconductor/metal Schottky junctions or other $MoTe_2/MoS_2$ heterojunctions. Hence, in the following, the overlapped area is considered as the effective photoactive area of the dual-floating vdWHs photodiode.

Benefitting from the strong built-in electric field of type II heterojunctions, the photoresponse speed of the dual-floating vdWHs photodiode is pretty fast. Figure 2e shows the time-resolved photocurrent of Device B that stimulated by a switchable 532 nm laser ($P_{in} = 42.4$ mW/cm$^2$) at frequencies of 1 Hz and 10 kHz. Despite a deviation from a well-defined square shape, the on-off behavior of the photoresponse remains apparent at 10 kHz, highlighting the potential in high-frequency photodetection. Measured between 10% and 90% of the maximum photocurrent, the rise time and decay time are both estimated as ~30 μs. A series of measurements conducted at different laser modulation frequencies (Figure 2f) show that the 3 dB bandwidth of the dual-floating vdWHs photodiode is ~10 kHz, which is in good agreement with the extracted photoresponse time.

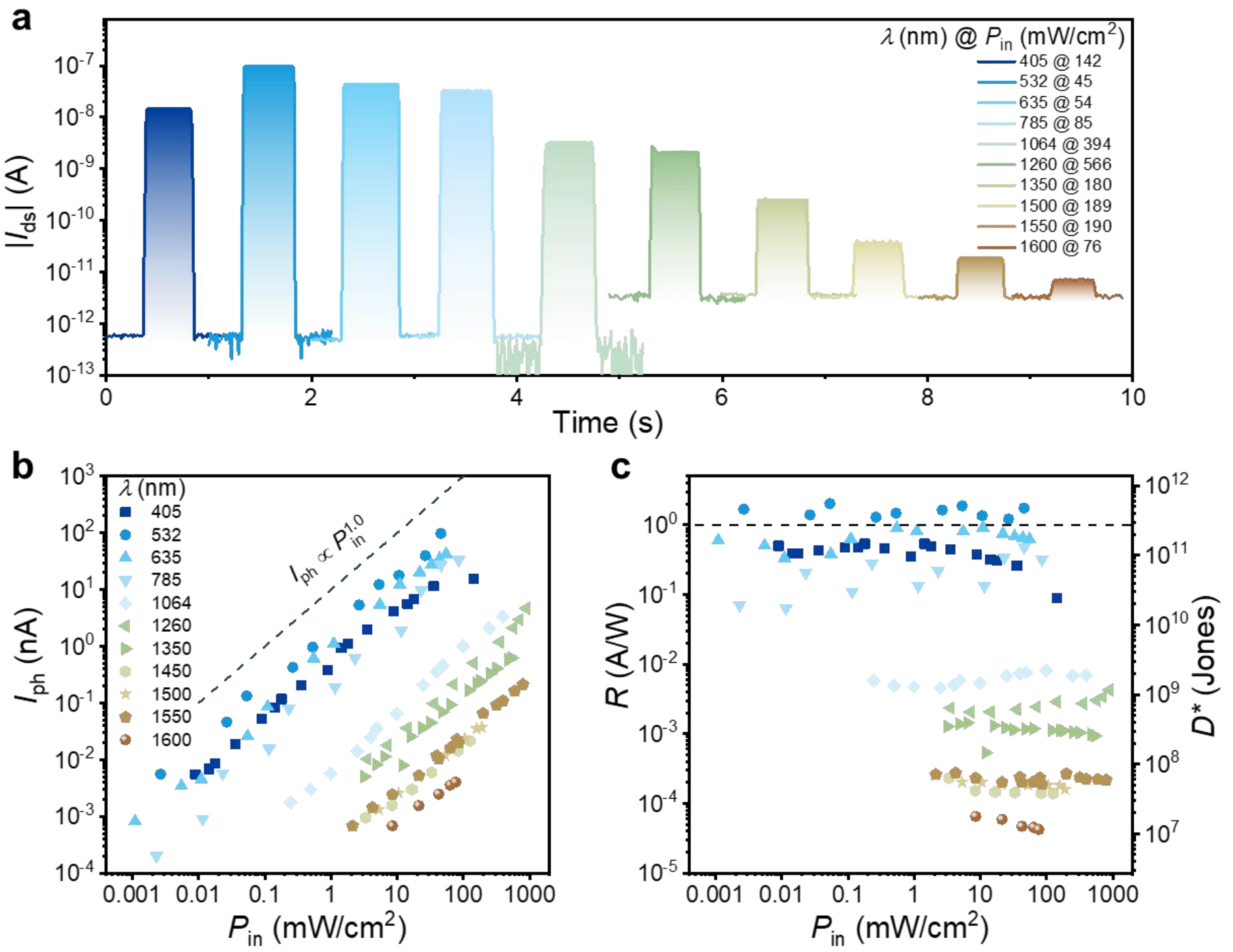

**Figure 3. Broadband photoresponse of the dual-floating vdWHs photodiode with excellent linearity. a**) Time-resolved photoresponse of Device B at the laser wavelength from 405 to 1600 nm. **b**), **c**) Photocurrent $I_{ph}$, responsivity $R$, and detectivity $D^*$ versus laser power density $P_{in}$ from visible to NIR bands. Here, $V_g = -50$ V and the measurements for 1260, 1350, 1500, 1550, and 1600 nm are carried out in the air.

As aforementioned, the photoresponse of $MoTe_2/MoS_2$ vdWH can be extended to the NIR band through interlayer excitation, beyond the intrinsic response bands of both $MoTe_2$ and $MoS_2$ layers. Here, we investigated the broadband detection capabilities of Device B from visible to NIR wavelength (405-1600 nm). Figures S7 and S8 show the results based on output curves for 635, 785, and 1064 nm, exhibiting similar performance charateristics to that for 532 nm. For better visualization, the transient photocurrent was recorded at zero drain-source bias while lasers with different wavelengths were switched with the modulation frequency of 1 Hz. The corresponding square photocurrent was generated for all chosen wavelength, with the dark current at picoampere level and the light on/off ratio upto $10^5$ for 532 nm, as shown in Figure 3a. The evident photoresponse at 1260-1600 nm is definitely beyond the bandgaps of both $MoTe_2$ and $MoS_2$ layers, thus confirming the validity of interlayer excitation in the dual-floating vdWHs. It should be noted that the measurements conducted under 1260-1600 nm illumination were performed at ambient conditions, and the observed increase in dark current can likely be attributed to the presence of absorbed oxygen or water molecules[41,42]. As anticipated, the extracted photocurrent is linearly dependent on the laser power density from $10^{-3}$ to $10^2$ mW cm$^{-2}$ (Figure 3b).

Generally, responsivity $R$ (input-output efficiency for a photodetector, in units of A/W):

$$R = \frac{I_{\mathrm{ph}}}{P_{\mathrm{in}}A}, \quad (3)$$

and specific detectivity $D^*$ (normalized signal-to-noise ratio of a photodetector on active area and bandwidth, in units of cm·Hz$^{1/2}$/W or Jones) [16]:

$$D^* = R\sqrt{\frac{A\Delta f}{\int_0^{\Delta f} S df}}, \quad (4)$$

are adopted to evaluate the photodetection performance, where $S$ is noise spectral density and $\Delta f$ is electrical bandwidth. Experimental results from 405 to 1600 nm are presented in Figures 3c and S9. In the linear region, $R$ is nearly constant for all measured wavelength. The average value of $R$~1.57 A/W is obtained at 532 nm, even larger than the maximum $R$~0.86 A/W of an ideal photodiode. This anomaly can be attributed to the opposite charge accumulation of two floating gates under illumination, resulting in an equivalent reverse bias applid on the photodiode[23], thereby significantly enhancing the photocurrent by times. The degree of enhancement can be quantified through the external quantum efficiency (ratio of photogenerated carriers to incident photons, $EQE = R \cdot hc/e\lambda$), where $h$ is Plank's constant, $c$ and $\lambda$ are light velocity and wavelength, respectively. Based on the experimental data, the $EQE$ is calculated to be ~365.8% for $\lambda = 532$ nm, which is several times larger than that of $WSe_2/MoS_2/WSe_2$[23] and Perovskite/BP/$MoS_2$[43] single-floating photodiodes. The $P_{\mathrm{in}}$ -independent $EQE$ confirms the negligible carrier trapping within the device[44]. $R$ decreases with the increase of $\lambda$ (Figure S9c), mainly ascribed to the weakened light absorption for longer wavelength. The specific detectivity follows the same tend as the responsivity. Under the Fourier transform of dark current, the noise spectral density is evaluated as the constant value of ~ $10^{-29}$ A$^2$/Hz (Figures S9a and S9b), and the average specific detectivity of ~ $4.28\times10^{11}$ Jones at $\lambda = 532$ nm is comparable to the state-of-art detectivities of other vdWHs photodiodes (Table S2). Furthermore, based on the decay time extracted from high-frequency measurements, a sudden rise is observed for $\lambda \geq 1260$ nm (see Figure S9d), indicating a slower photoresponse process for interlayer excitation than for intralayer excitation.

Based on the electrical characterization depicted in Figure S4, the dual-floating vdWHs photodiode exhibits reconfigurability into *p-n* type (regions II and III) and $n^+$-*n* type (region IV) when the gate voltage is varied from –60 to 60 V. To explore the impact of gate voltage (or electric configuration) on the linearity, we conducted photocurrent measurements under different gate voltages with Device C, as shown in Figures 4 and S10. Encouragingly, the device maintains excellent linearity which is nearly independent of the gate voltage, as evidenced in Figure S10b. The only notable difference brought about by the gate voltage is the photocurrent magnitude. Figure 4b illustrates that the photocurrent generated in the *p-n* configuration is one order of magnitude larger than that in the $n^+$-*n* configuration. The probable reason can be inferred from the band structure. It is important to note that the direction of the photocurrent remains consistent, suggesting the unchanged type II heterojunctions of the vdWHs in both *p-n* and $n^+$-*n* configurations. Therefore, the band diagrams can be illustrated as shown in Figures 4c and 4d. As the gate voltage increases (from *p-n* to $n^+$-*n* configuration), the band bending gradually alleviates, resulting in a weaker built-in electric field and reduced efficiency in separating photogenerated electron-hole pairs. Moreover, the transmission efficiency for holes in the vdWHs is also diminished due to the rising Fermi level. Consequently, the photocurrent becomes smaller in the $n^+$-*n* configuration.

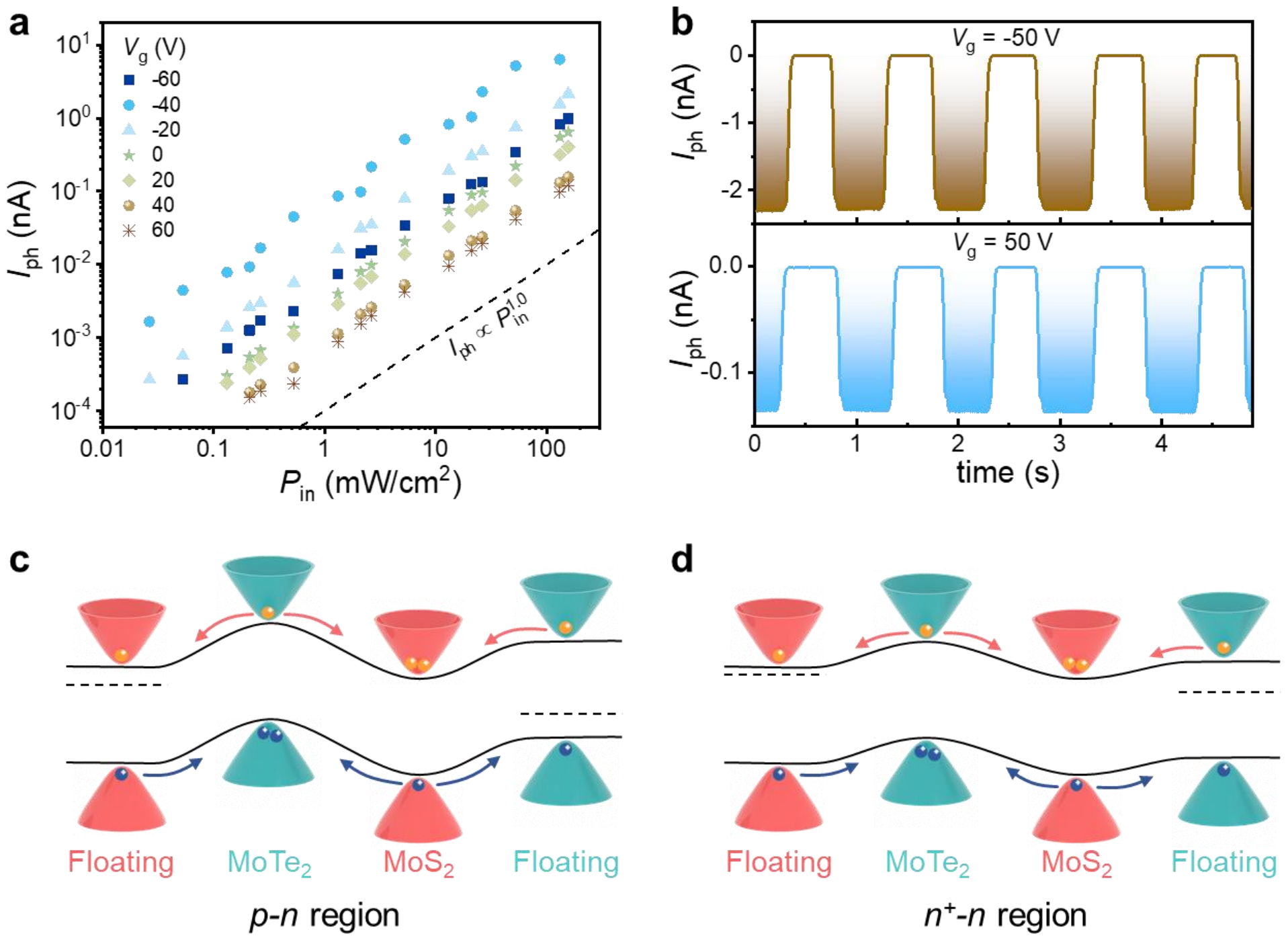

**Figure 4. Gating-independent linearity of the dual-floating vdWHs photodiode. a)** Excellent linearity of Device C (with thicknesses of 6/9/11/26 nm for $MoS_2$/$MoTe_2$/$MoS_2$/$MoTe_2$) holds in the gate voltage range from -60 to 60 V, where the laser wavelength is 635 nm. **b)** The photocurrent at *p-n* region ($V_g = -50$ V) and $n^+$-*n* region

($V_g = 50$ V), respectively, when $P_{in} = 157$ mW/cm$^{-2}$. **c**) and **d**) illustrated the band diagram and the photocarrier dynamics at *p-n* region and $n^+$-*n* region, respectively. The dashed lines denote the Fermi levels.

### 2.4. Potential applications and the configuration universality

In accordance with the band alignment, the photoresponse of the dual-floating vdWHs photodiode is also contingent on the gate voltage, because of the gate-tunable Fermi level in vdWHs. As demonstrated in Figure S11a, a large $V_{oc}$ is achieved with Device B in the $V_g$ range from –70 to –30 V, while it reduces to a low level when $V_g <- 80$ V or $V_g > 80$ V. After properly defined, the device can operate as an optoelectronic AND logic gate, as shown in Figures S11b and S11c. Besides, the device exhibits potential for high-performance imaging attributed to its excellent performance of fast, broadband, and linear photoresponse. Figure 5 showcases a simulated imaging system with a single pixel and its corresponding imaging outcomes. The pattern on aluminum mask can be reconstituted by the photocurrent through the program-controlled 2D motorized stage. The images of English letters "IFE" and Chinese character of "Flexible" have been obtained using four distinct wavelengths of 405, 532, 785, and 1064 nm (Figures 5b and 5c), validating the stable and panchromatic imaging function of our photodiode. Specifically, the excellent linearity contributes to achieving high-resolution imaging.

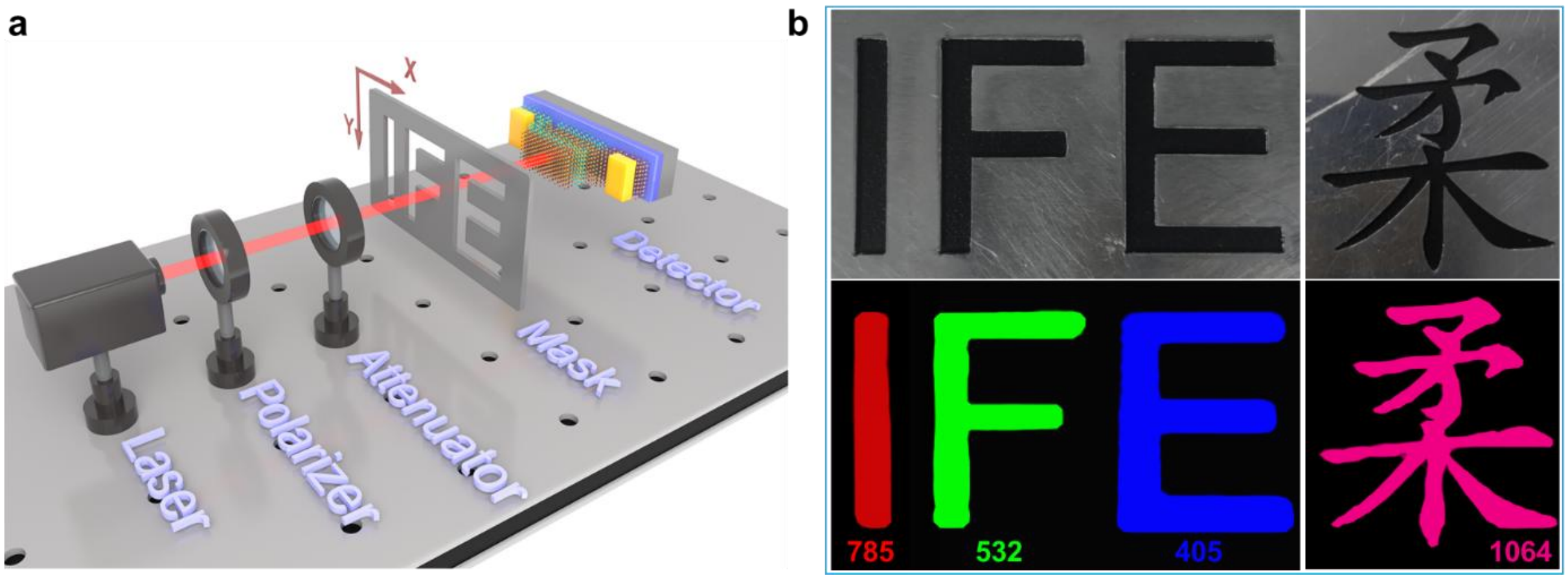

**Figure 5. Multiband imaging capacity with the dual-floating vdWHs photodiode. a**) Schematic of a home-built mimic imaging system for single-pixel device. **b**) Obtained images of letters "IFE" and the Chinese character of "Flexible" using Device B under 405, 532, 785, and 1064 nm illumination, respectively, when $V_g =- 50$ V. Upper panels: aluminum masks, lower panels: photocurrent images.

To explore the universality of the dual-floating vdWHs photodiode configuration, we have fabricated the device once more using ambipolar $WSe_2$ layers and $n$-type $MoS_2$ layers (Figure S12a). Compared to the $MoS_2/MoTe_2/MoS_2/MoTe_2$ dual-floating device, the $MoS_2/WSe_2/MoS_2/WSe_2$ dual-floating photodiode displays a larger rectification ratio and a lower dark current. Under the 532 nm illumination at $P_{\text{in}} = 43$ mW/cm$^2$, the maximum $V_{\text{oc}}$ and $I_{\text{sc}}$ reach 0.36 V and 19 nA, respectively, as shown in Figure S12b. Similar trend on $P_{\text{in}}$ was also observed for $PCE$, $I_{\text{sc}}$, $V_{\text{oc}}$, $I_{\text{ph}}$, and $R$ (see Figures S12c-S12f), while the maximum value of responsivity was obtained at 635 nm (i.e., $R \sim 1.0$ A/W). As anticipated, the device exhibits exceptional linearity (with $LDR \sim 100$ dB) at all tested wavelengths (including 405, 532, 635, and 785 nm).

## 3. Conclusion

In summary, dual-floating vdWHs photodiodes with robust linear photoresponse have been demonstrated by vdWHs stacking a top $MoS_2$ floating layer, the middle $MoTe_2/MoS_2$ conducting channel and a bottom $MoTe_2$ floating layer. Benefited from the ambipolar $MoTe_2$ and $n$-type $MoS_2$, two extra type II heterojunctions are formed at either side of the middle $MoTe_2/MoS_2$ conducting channel. With the assistance of three type II heterojunctions, the carriers in channel layers are effectively depleted or transferred out of there, resulting in minimal carrier interaction (such as carrier trapping in channel layers), and leading to robust linear photoresponse even in non-ideal photodiodes. In addition, the photoresponse of the dual-floating vdWHs photodiode can be extended to the NIR band by interlayer excitation, beyond the bands determined by intralayer excitation in both $MoTe_2$ and $MoS_2$ layers. The experimental results show that this device performs a broadband and self-driven photodetection, ranging from visible to NIR bands (from 405 to 1600 nm), with the persistent linear photoresponse for various gate voltages. For 532 nm, a high $R$ of ~ 1.57 A/W and a high $D^*$ of ~ $4.28 \times 10^{11}$ Jones with fast response (~ 30 μs) are obtained, comparable to the state-of-the-art performances of 2D material photodiodes. In short, the dual-floating vdWHs photodiode configuration with linear photoresponse presents a promising device concept for fast detection, high-resolution imaging, and logic optoelectronics.

## 4. Experimental Section

*Device fabrication*: All the multilayer $MoTe_2$, $MoS_2$, and hBN nanosheets were exfoliated from commercially available bulk crystals (HQ Graphene Company) by micromechanical cleavage technique with Scotch-tape. The dual-floating vdWHs photodiode was then fabricated by the dry transfer technique with the assistance of polydimethylsiloxane (PDMS) stamp. Firstly, hBN nanosheet was deposited onto the pre-cleaned $SiO_2/Si$ substrate (highly $n$- doped Si with resistivity < 0.01 Ω cm, the $SiO_2$ thickness ~ 285 nm). Bottom floating $MoTe_2$ nanosheet was then transferred in sequence on the hBN. Secondly, a $MoS_2$ nanosheet was stacked on the bottom floating $MoTe_2$, and another $MoTe_2$ was transferred on the $MoS_2$ to form a $MoTe_2/MoS_2$ junction channel. Note that, the bottom floating $MoTe_2$ was covered by $MoS_2$, and the channel $MoTe_2$ was not in contact with the bottom floating $MoTe_2$. Thirdly, another $MoS_2$ was stacked on the channel $MoTe_2$. Similarly, the top floating $MoS_2$ was not in contact with the channel $MoS_2$. Finally, Au (thickness ~ 50 nm) electrodes were transferred onto the $MoTe_2$ and $MoS_2$ channels as drain and source electrodes, respectively. The as-prepared device was annealed at 200 °C in an $Ar/H_2$ (90%/10%) atmosphere tube furnace for 0.5 hours to remove the resisted residues and to decrease contact resistance. The fabrication process is also shown in Figure S1.

*Device fabrication*: Raman and PL spectra of $MoTe_2$, $MoS_2$ and their heterostructure were collected by using a confocal mirco-Raman system (Alpha300R, WITec) excited by 532 nm laser (spot size ~ 400 nm, laser power ~ 1.0 mW, and resolution ~ 0.02 $cm^{-1}$). AFM and KPFM images were obtained by an atomic force microscope (Dimension Icon, Bruker). All electrical/optoelectrical measurements were carried out at room temperature by a semiconductor parameter analyzer (PDA FS380 Pro, Platform Design Automation) in a probe station with a vacuum degree of ~$5\times10^{-6}$ mbar, except the NIR photoresponse behavior that tested in the air. The light sources were lasers with wavelengths of 405, 532, 635, 785, 1064, 1260, 1350, 1450, 1500, 1550 and 1600 nm, respectively.

**Supporting Information**

Supporting Information is available from the Wiley Online Library or from the author.

**Acknowledgements**

J.Xu, X.Luo, and X.Lin contributed equally to this work. This work was supported by the National Natural Science Foundation of China (Grant Nos. 61905198 and 62274087), the National Postdoctoral Program for Innovative Talents (No. BX20190283), and the Fundamental Research Funds for the Central Universities. Authors also thank the Analytical & Testing Center of NPU for the assistance in device fabrication.

References

[1] T. Mueller, F. Xia, P. Avouris, *Nat. Photonics* **2010**, 4, 297.

[2] W. Bogaerts, D. Pérez, J. Capmany, D. A. B. Miller, J. Poon, D. Englund, F. Morichetti, A. Melloni, *Nature* **2020**, 586, 207.

[3] L. Mennel, J. Symonowicz, S. Wachter, D. K. Polyushkin, A. J. Molina-Mendoza, T. Mueller, *Nature* **2020**, 579, 62.

[4] A. Dodda, D. Jayachandran, A. Pannone, N. Trainor, S. P. Stepanoff, M. A. Steves, S. S. Radhakrishnan, S. Bachu, C. W. Ordonez, J. R. Shallenberger, J. M. Redwing, K. L. Knappenberger, D. E. Wolfe, S. Das, *Nat. Mater.* **2022**, 21, 1379.

[5] X. Gan, R.-J. Shiue, Y. Gao, I. Meric, T. F. Heinz, K. Shepard, J. Hone, S. Assefa, D. Englund, *Nat. Photonics* **2013**, 7, 883.

[6] F. H. L. Koppens, T. Mueller, P. Avouris, A. C. Ferrari, M. S. Vitiello, M. Polini, *Nat. Nanotechnol.* **2014**, 9, 780.

[7] F. P. García de Arquer, A. Armin, P. Meredith, E. H. Sargent, *Nat. Rev. Mater.* **2017**, 2, 16100.

[8] G. Konstantatos, E. H. Sargent, *Nat. Nanotechnol.* **2010**, 5, 391.

[9] M. Long, P. Wang, H. Fang, W. Hu, *Adv. Funct. Mater.* **2019**, 29, 1803807.

[10] N. J. Huo, G. Konstantatos, *Adv. Mater.* **2018**, 30, 1801164.

[11] Y. Liu, N. O. Weiss, X. Duan, H.-C. Cheng, Y. Huang, X. Duan, *Nat. Rev. Mater.* **2016**, 1, 16042.

[12] K. Zhang, T. Zhang, G. Cheng, T. Li, S. Wang, W. Wei, X. Zhou, W. Yu, Y. Sun, P. Wang, D. Zhang, C. Zeng, X. Wang, W. Hu, H. J. Fan, G. Shen, X. Chen, X. Duan, K. Chang, N. Dai, *ACS Nano* **2016**, 10, 3852.

[13] C.-H. Lee, G.-H. Lee, A. M. van der Zande, W. Chen, Y. Li, M. Han, X. Cui, G. Arefe, C. Nuckolls, T. F. Heinz, J. Guo, J. Hone, P. Kim, *Nat. Nanotechnol.* **2014**, 9, 676.

[14] O. Lopez-Sanchez, D. Lembke, M. Kayci, A. Radenovic, A. Kis, *Nat. Nanotechnol.* **2013**, 8, 497.

[15] J. Xu, X. Luo, S. Hu, X. Zhang, D. Mei, F. Liu, N. Han, D. Liu, X. Gan, Y. Cheng, W. Huang, *Adv. Mater.* **2021**, 33, 2008080.

[16] F. Liu, J. Xu, Y. Yan, J. Shi, S. Ahmad, X. Gan, Y. Cheng, X. Luo, *ACS Photonics* **2023**, 10, 1126.

[17] S. Hu, X. Luo, J. Xu, Q. Zhao, Y. Cheng, T. Wang, W. Jie, A. Castellanos-Gomez, X. Gan, J. Zhao, *Adv. Electron. Mater.* **2022**, 8, 2101176.

[18] S. Hu, J. Xu, Q. Zhao, X. Luo, X. Zhang, T. Wang, W. Jie, Y. Cheng, R. Frisenda, A. Castellanos‐Gomez, X. Gan, *Adv. Opt. Mater.* **2021**, 9, 2001802.

[19] G. Wu, X. Wang, Y. Chen, S. Wu, B. Wu, Y. Jiang, H. Shen, T. Lin, Q. Liu, X. Wang, P. Zhou, S. Zhang, W. Hu, X. Meng, J. Chu, J. Wang, *Adv. Mater.* **2020**, 32, 1907937.

[20] F. Liu, Y. Yan, D. Miao, J. Xu, J. Shi, X. Gan, Y. Cheng, X. Luo, *Appl. Surf. Sci.* **2023**, 616, 156444.

[21] Y. Chen, X. Wang, G. Wu, Z. Wang, H. Fang, T. Lin, S. Sun, H. Shen, W. Hu, J. Wang, J. Sun, X. Meng, J. Chu, *Small* **2018**, 14, 1703293.

[22] W. Choi, I. Akhtar, D. Kang, Y.-j. Lee, J. Jung, Y. H. Kim, C.-H. Lee, D. J. Hwang, Y. Seo, *Nano Lett.* **2020**, 20, 1934.

[23] Y. Jiang, R. Wang, X. Li, Z. Ma, L. Li, J. Su, Y. Yan, X. Song, C. Xia, *ACS Nano* **2021**, 15, 14295.

[24] A. Castellanos-Gomez, M. Buscema, R. Molenaar, V. Singh, L. Janssen, H. S. J. van der Zant, G. A. Steele, *2D Mater.* **2014**, 1, 011002.

[25] Y. Liu, J. Guo, E. Zhu, L. Liao, S.-J. Lee, M. Ding, I. Shakir, V. Gambin, Y. Huang, X. Duan, *Nature* **2018**, 557, 696.

[26] H. Li, Q. Zhang, C. C. R. Yap, B. K. Tay, T. H. T. Edwin, A. Olivier, D. Baillargeat, *Adv. Funct. Mater.* **2012**, 22, 1385.

[27] C. Ruppert, O. B. Aslan, T. F. Heinz, *Nano Lett.* **2014**, 14, 6231.

[28] A. Splendiani, L. Sun, Y. Zhang, T. Li, J. Kim, C.-Y. Chim, G. Galli, F. Wang, *Nano Lett.* **2010**, 10, 1271.

[29] X. Zhang, H. Nan, S. Xiao, X. Wan, Z. Ni, X. Gu, K. Ostrikov, *ACS Appl. Mater. Interfaces* **2017**, 9, 42121.

[30] K. F. Mak, K. He, C. Lee, G. H. Lee, J. Hone, T. F. Heinz, J. Shan, *Nat. Mater.* **2013**, 12, 207.

[31] X. Wang, P. Wang, J. Wang, W. Hu, X. Zhou, N. Guo, H. Huang, S. Sun, H. Shen, T. Lin, M. Tang, L. Liao, A. Jiang, J. Sun, X. Meng, X. Chen, W. Lu, J. Chu, *Adv. Mater.* **2015**, 27, 6575.

[32] S. Schmitt-Rink, D. S. Chemla, D. A. B. Miller, *Phys. Rev. B* **1985**, 32, 6601.

[33] L. Yin, F. Wang, R. Cheng, Z. Wang, J. Chu, Y. Wen, J. He, *Adv. Funct. Mater.* **2019**, 29, 1804897.

[34] N. T. Duong, J. Lee, S. Bang, C. Park, S. C. Lim, M. S. Jeong, *ACS Nano* **2019**, 13, 4478.

[35] W. Ahmad, L. Pan, K. Khan, L. P. Jia, Q. D. Zhuang, Z. M. Wang, *Adv. Funct. Mater.* **2023**, 33, 202300686.

[36] E. Wu, Y. Xie, Q. Liu, X. Hu, J. Liu, D. Zhang, C. Zhou, *ACS Nano* **2019**, 13, 5430.

[37] Y. Pan, X. Liu, J. Yang, W. J. Yoo, J. Sun, *ACS Appl. Mater. Interfaces* **2021**, 13, 54294.

[38] X. Liu, D. Qu, H.-M. Li, I. Moon, F. Ahmed, C. Kim, M. Lee, Y. Choi, J. H. Cho, J. C. Hone, W. J. Yoo, *ACS Nano* **2017**, 11, 9143.

[39] X. Lu, L. Sun, P. Jiang, X. Bao, *Adv. Mater.* **2019**, 31, 1902044.

[40] F. Wu, Q. Li, P. Wang, H. Xia, Z. Wang, Y. Wang, M. Luo, L. Chen, F. Chen, J. Miao, X. Chen, W. Lu, C. Shan, A. Pan, X. Wu, W. Ren, D. Jariwala, W. Hu, *Nat. Commun.* **2019**, 10, 4663.

[41] H. Qiu, L. Pan, Z. Yao, J. Li, Y. Shi, X. Wang, *Appl. Phys. Lett.* **2012**, 100, 123104.

[42] B. Radisavljevic, A. Radenovic, J. Brivio, V. Giacometti, A. Kis, *Nat. Nanotechnol.* **2011**, 6, 147.

[43] L. Wang, X. Zou, J. Lin, J. Jiang, Y. Liu, X. Liu, X. Zhao, Y. F. Liu, J. C. Ho, L. Liao, *ACS Nano* **2019**, 13, 4804.

[44] D. Kufer, G. Konstantatos, *Nano Lett.* **2015**, 15, 7307.

# Supporting Information

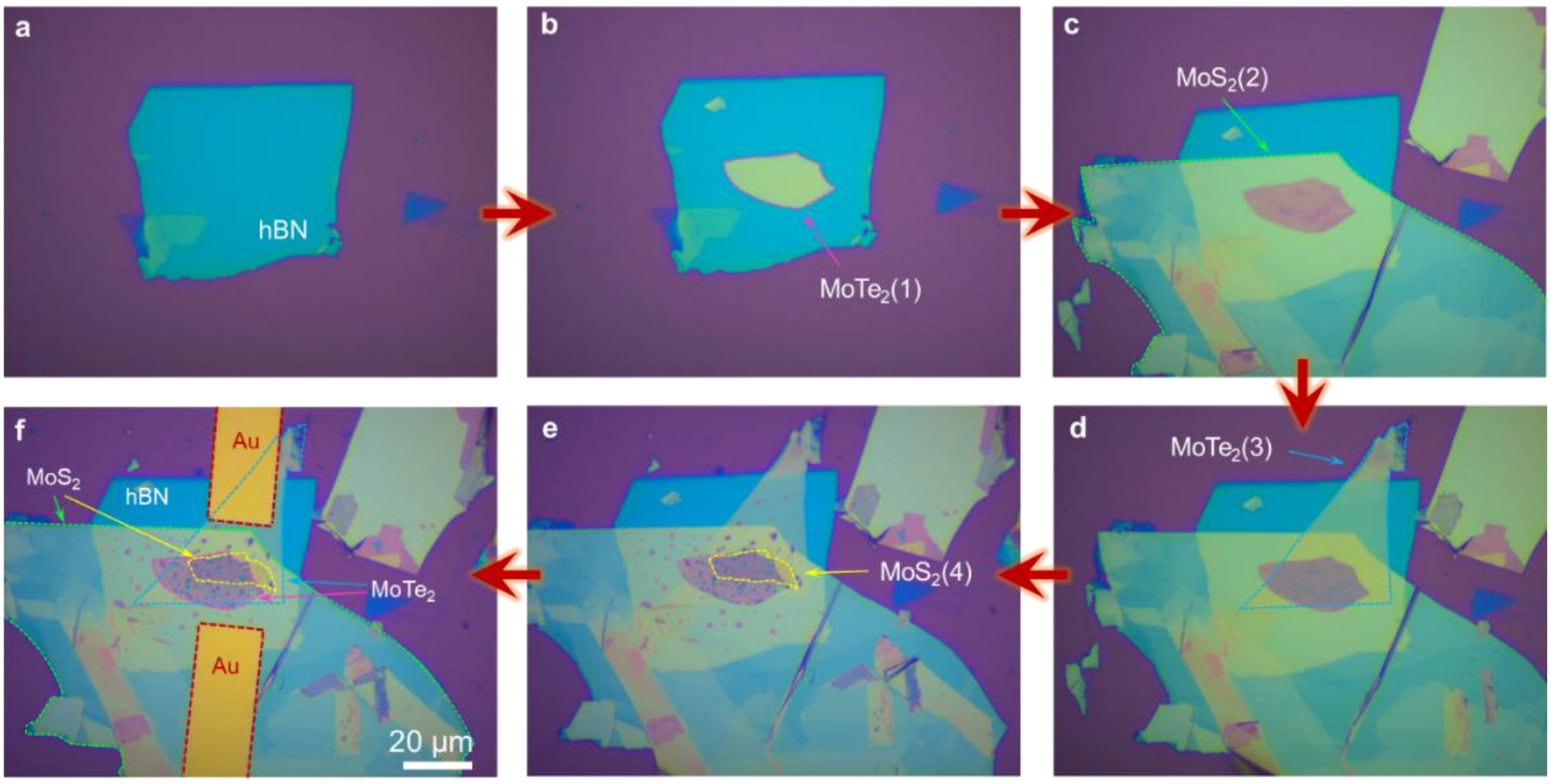

**Figure S1** Device fabrication process. **a** Exfoliated hBN was transferred on the pre-cleaned $SiO_2$/Si substrate. **b** $MoTe_2$ bottom floating layer was then transferred on the hBN. **c** $MoS_2$ was stacked on the bottom floating layer and covered the bottom floating layer completely. **d** Another $MoTe_2$ was transferred on the $MoS_2$ to form a junction channel. **e** $MoS_2$ top floating layer was stacked on the $MoTe_2$ channel. **f** Two Au films (thickness ~ 50 nm) were transferred onto the $MoTe_2$ and the $MoS_2$ channel as the drain and the source electrode, respectively. Based on the optical image, the average width and length of this device (Device B) are ~ 19 and ~ 6.5 μm, respectively.

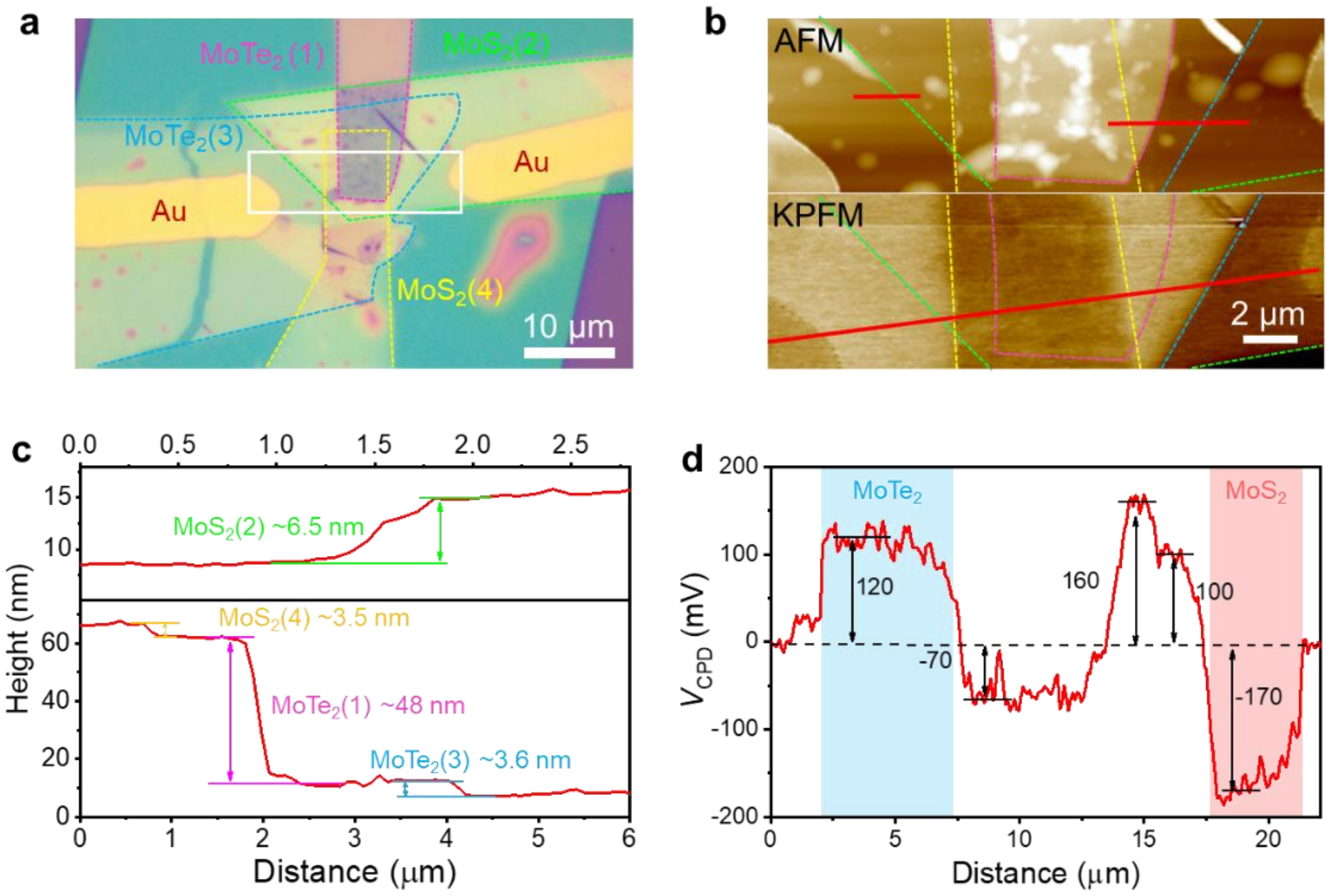

**Figure S2 a** Optical image of Device A. **b** AFM and KPFM mapping of the region in the white box of **a**. **c** Height profile along the red sold lines in the AFM image. The thicknesses of nanosheets from top to bottom can be identified: $MoS_2$ (4), $MoTe_2$ (3), $MoS_2$ (2), and $MoTe_2$ (1), are about 3.5, 3.6, 6.5, and 48 nm, respectively. **d** Contact potential difference profile along the red sold line in the KPFM image, the colored boxes highlight the homogeneous $MoTe_2$ layer and homogeneous $MoS_2$ layer, respectively.

KPFM mode is employed to measure the surface work function of the vdWHs by the contact potential difference (CPD) between the sample surface and AFM probe tip. The CPD between the tip and the sample is defined as follows: $V_{\text{CPD}} = (W_{\text{tip}} - W_{\text{sample}})/e$, where $W_{\text{tip}}$ and $W_{\text{sample}}$ are the work functions of the tip and the sample, respectively, and $e$ is the elementary charge. We use the Au electrode as the standard reference to calculate the work function of each layer, i.e., $V_{\text{Au-sample}} = (W_{\text{Au}} - W_{\text{sample}})/e$. Hence, the work function of $MoTe_2$ and $MoS_2$ can be written as

$$W_{\text{MoTe}_2} = W_{\text{Au}} - V_{\text{Au-MoTe}_2} \times e \quad \text{(S1)}$$

$$W_{\text{MoS}_2} = W_{\text{Au}} - V_{\text{Au-MoS}_2} \times e \quad \text{(S2)}$$

Combined with the experimental data of Figure S2d and $W_{\text{Au}}$~5.2 eV, the work function can be calculated and presented in Figure 1b. It is found that $MoTe_2$ has a higher work function (~5.32 eV) than Au, while $MoS_2$ is the lowest (~5.03 eV).

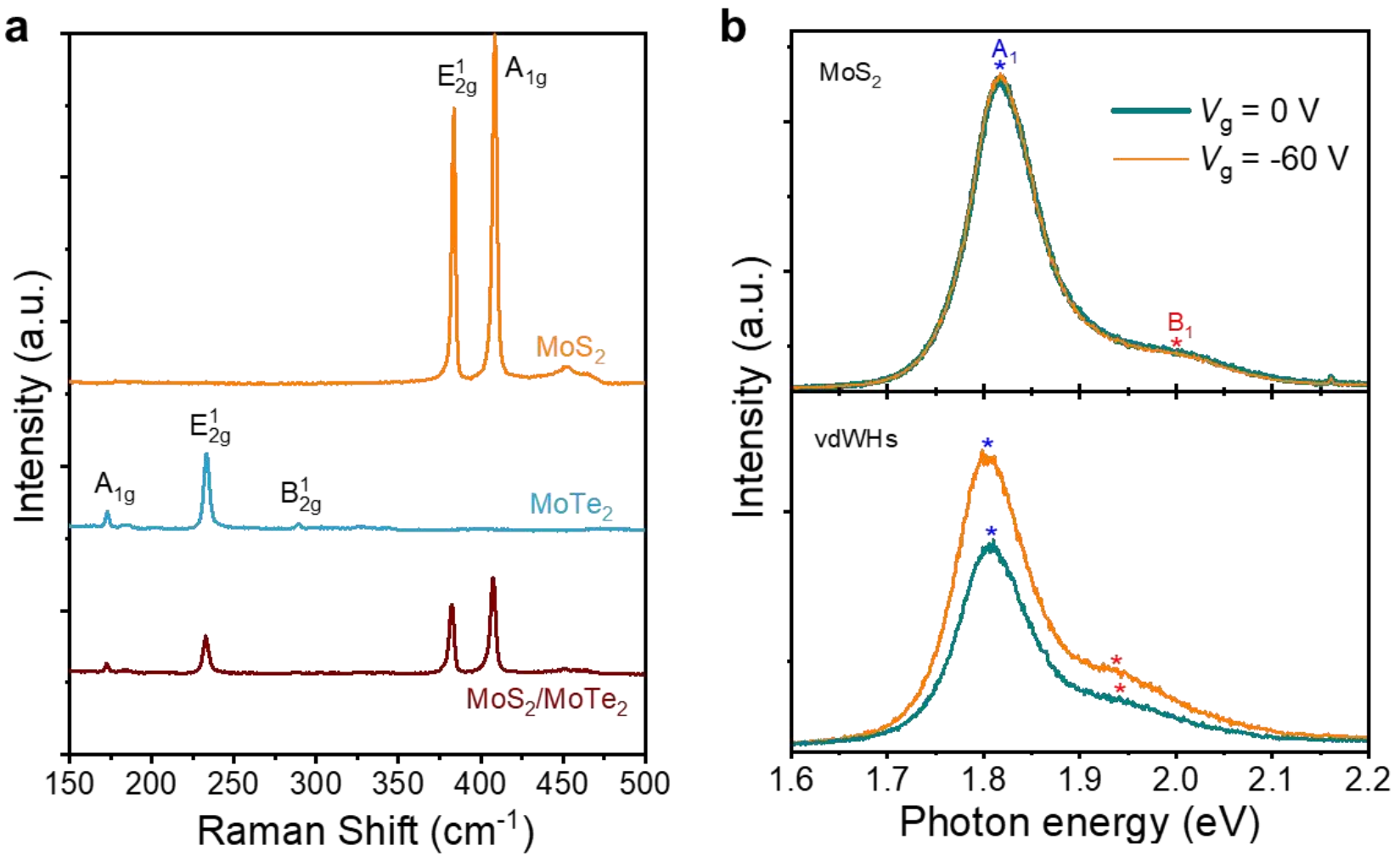

**Figure S3 a** Raman spectra of the homogeneous $MoS_2$ layer (orange), the homogeneous $MoTe_2$ layer (light blue) and the $MoS_2/MoTe_2/MoS_2/MoTe_2$ vdWHs (vermilion). Two intrinsic Raman peaks of $MoS_2$ at around 383 $cm^{-1}$ and 407 $cm^{-1}$ are assigned to the in-plane vibrational mode $E^1_{2g}$ and the out-of-plane vibrational mode $A_{1g}$, respectively[1]. For $MoTe_2$, these two modes emerge at around 233 $cm^{-1}$ and 172 $cm^{-1}$, respectively, and the out-of-plane bulk-inactive vibrational mode $B^1_{2g}$ appear at around 289.5 $cm^{-1}$, in agreement with the previous reports[2]. The characteristic Raman peaks of both $MoS_2$ and $MoTe_2$ appear again in the Raman spectrum of the vdWHs, indicating the high quality of the heterostructure. **b** PL spectra of the homogeneous $MoS_2$ layer (upper panel) and the $MoS_2/MoTe_2/MoS_2/MoTe_2$ vdWHs (lower panel) at $V_g = 0$ and -60 V, respectively. $A_1$ and $B_1$ peaks correspond to the direct excitonic transitions with the energy split from spin-orbital coupling of valence band. For the homogeneous $MoS_2$ layer, $A_1$ and $B_1$ peaks appear at ~1.817 eV and ~2.0 eV, respectively. For the vdWHs when $V_g$ varies from 0 to -60 V, the peak energy of $A_1$ decreases from ~1.807 to ~1.803 eV, and the peak energy of $B_1$ decreases from ~1.942 to ~1.936 eV.

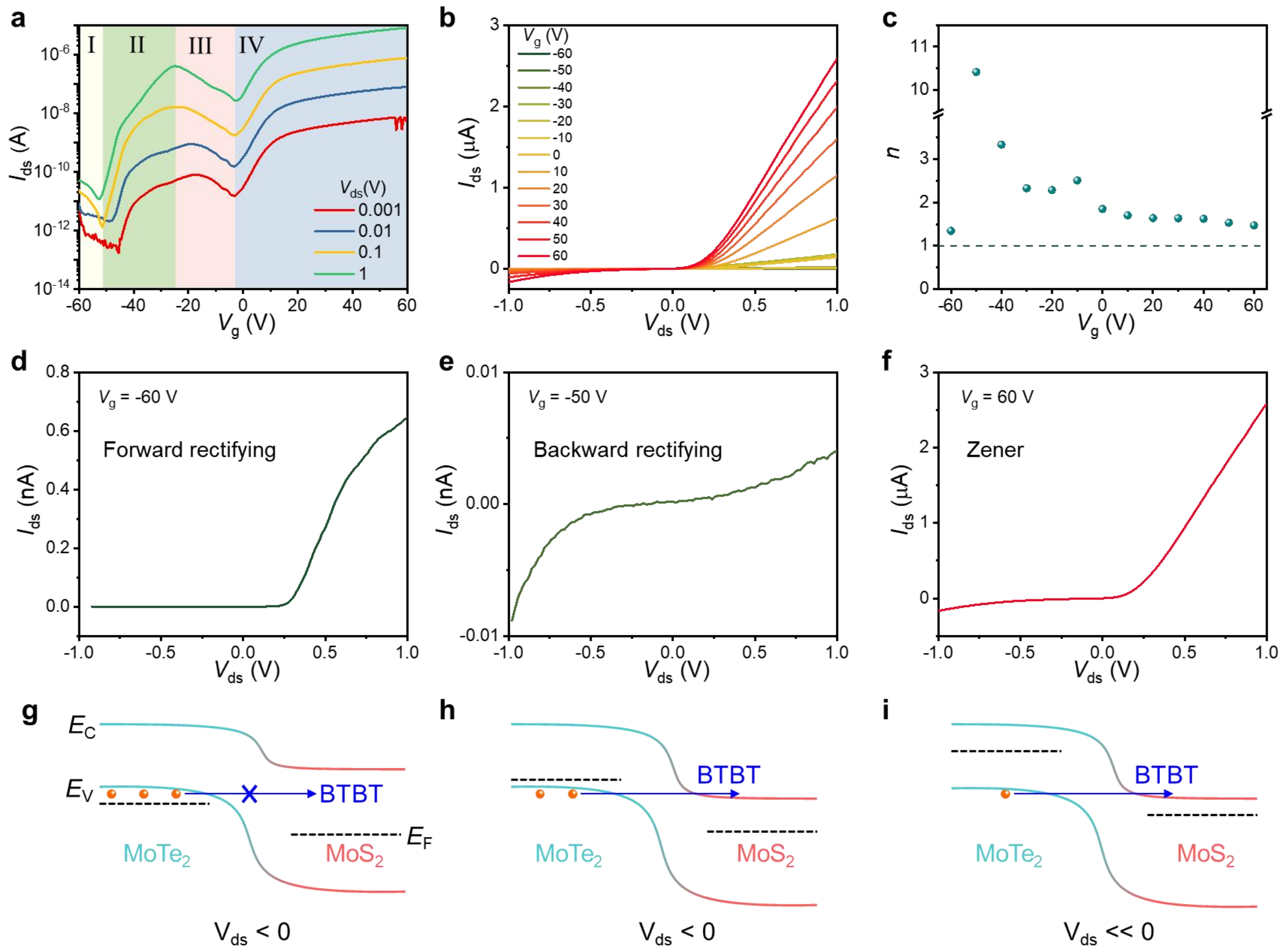

**Figure S4** Electrical characteristics of Device A. **a** Transfer curves at different $V_{ds}$. Restriction factors for electrical conduction: $MoS_2$ channel is limited (regime I), both $MoS_2$ and $MoTe_2$ channel limited (regime II), $MoTe_2$ channel limited (regime III), and barriers at the junction interfaces (regime IV). **b** Output curves at different $V_g$ under dark and 532 nm illumination condition. **c** Fitted ideal factor $n$ in the region of $V_{ds} > 0$ with the formula of $I_{ds} = I_S \left[\exp\left(\frac{eV_{ds}}{nkT}\right) - 1\right]$, where $I_S$ is the reverse saturation current. The device performs as: **d**, **g** the forward rectifying diode, **e**, **h** the backward rectifying diode, and **f**, **i** Zener diode, respectively, at the specific gate voltage. BTBT in the band diagrams denotes the band-to-band tunneling of electrons.

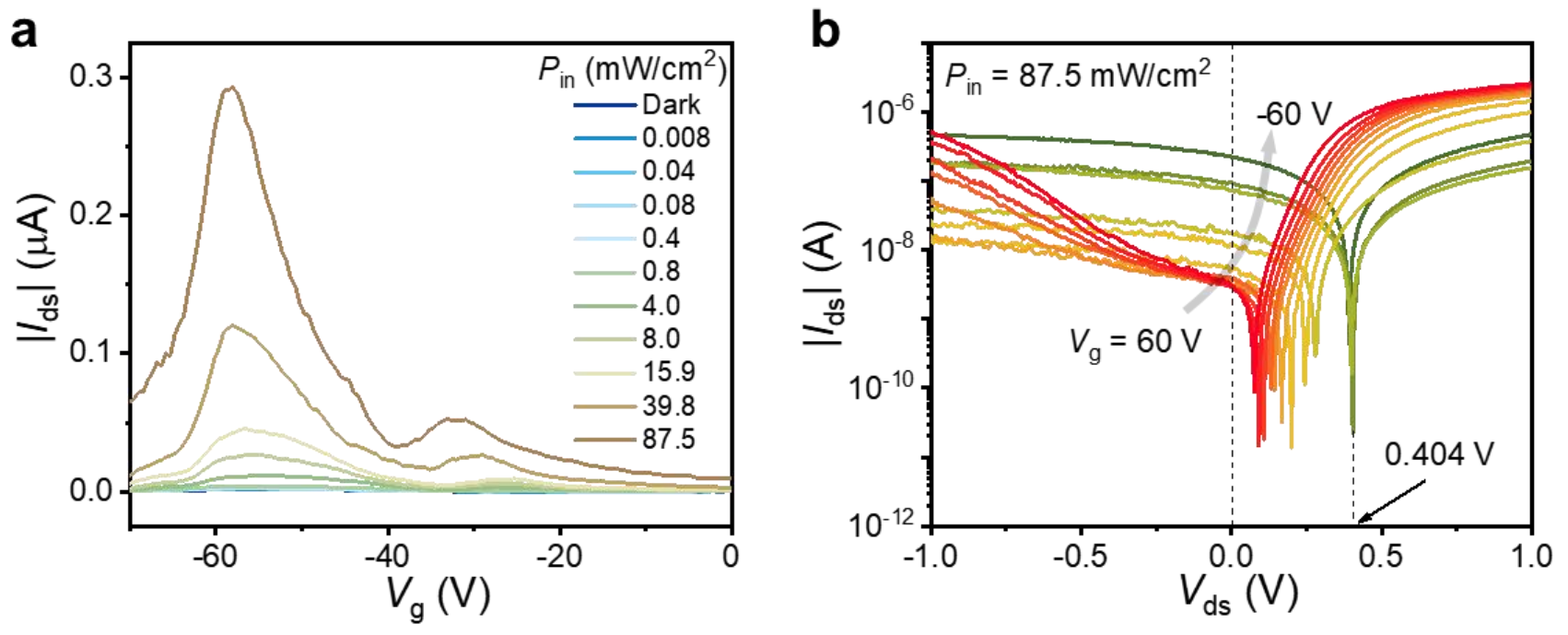

**Figure S5** **a** Transfer curves of Device A under 532 nm illumination with different laser power densities, here $V_{ds} = -0.1$ V is applied. **b** Output curves at different $V_g$ when the laser power density $P_{in} = 87.5$ mW/cm$^2$, where the maximum open-circuit voltage $V_{oc} = 0.404$ V is found at $V_g = -50$ V.

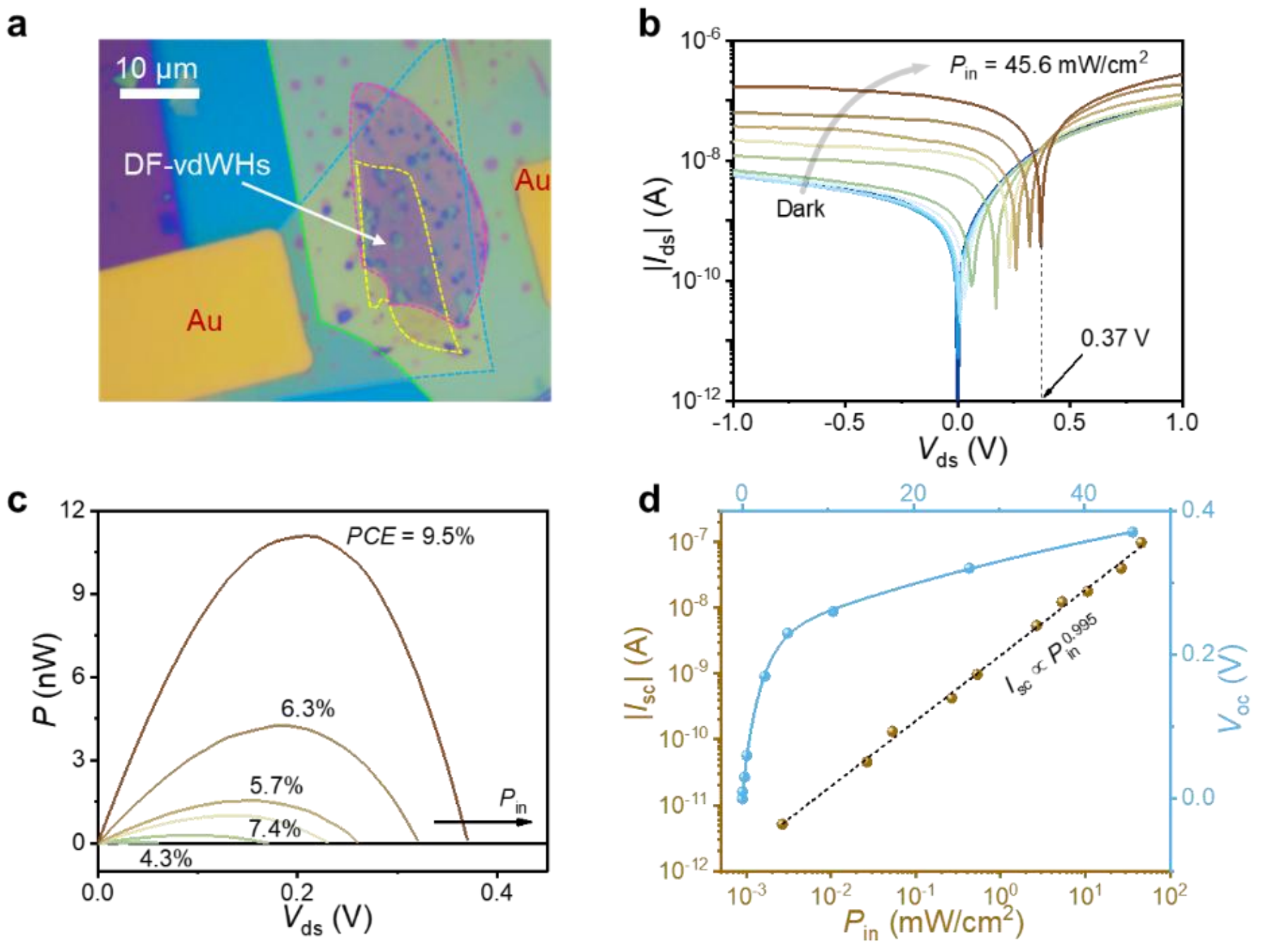

**Figure S6** Photoresponse characteristics of Device B under 532 nm illumination. **a** Optical image of the device. **b** Output curves under illumination with different intensities at $V_g = -50$ V. **c** Electrical power and PCE at different intensities. **d** Extracted $I_{sc}$ and $V_{oc}$ from the output curves.

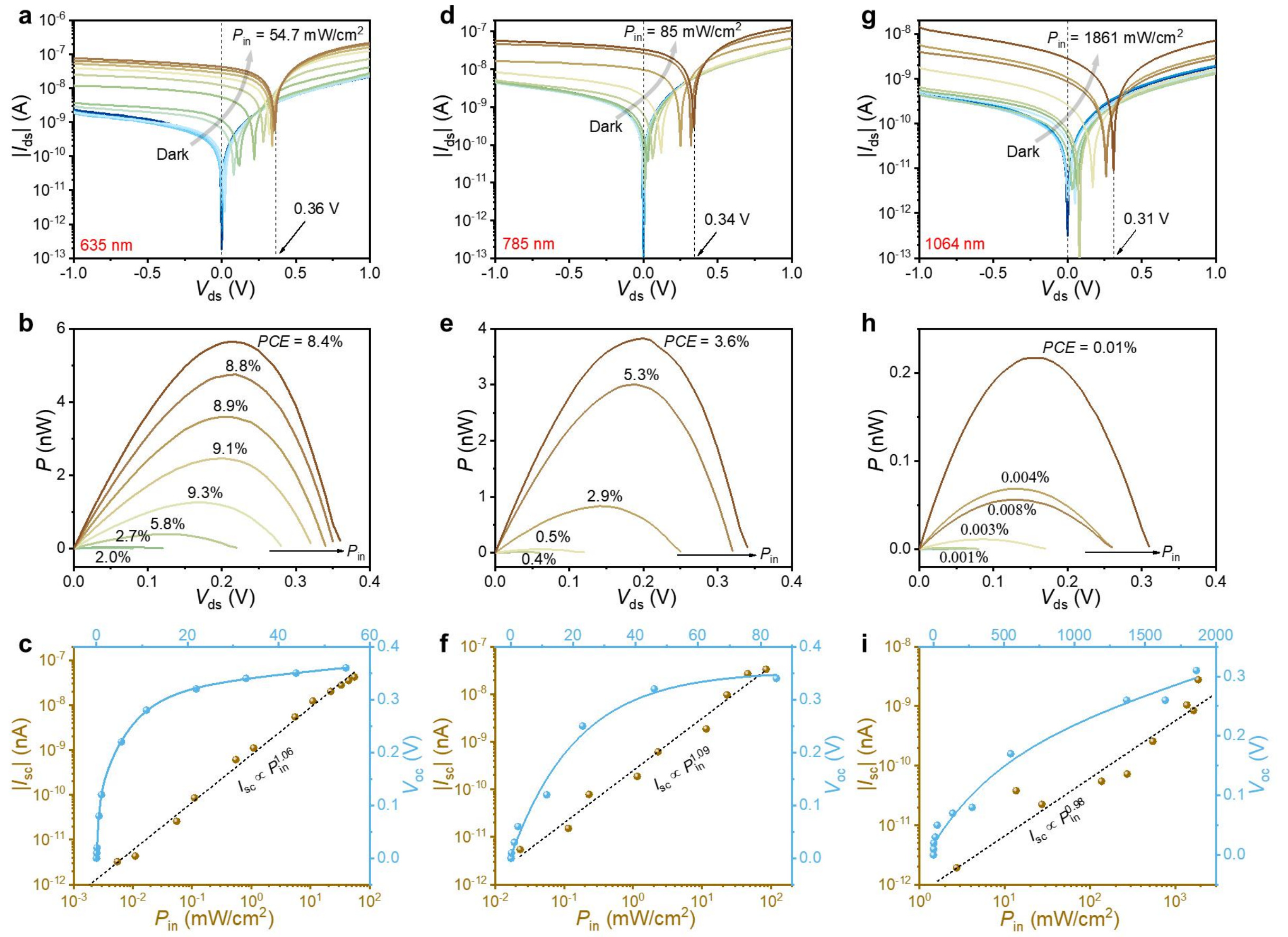

**Figure S7** Photoresponse characteristics of Device B for other laser wavelengths at $V_g = -50$ V: **a**-**c** 635 nm, **d**-**f** 785 nm, and **g**-**i** 1064 nm.

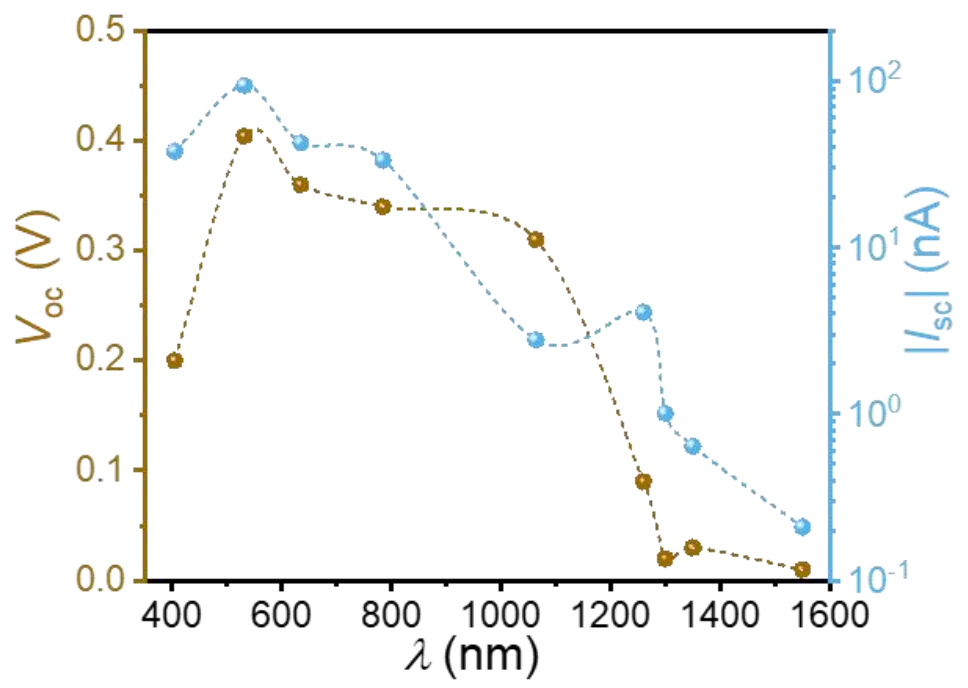

**Figure S8** The maximum $V_{oc}$ and $I_{sc}$ based on the output curves at different wavelength. It should be noted that these values are not saturated, and are extracted at different laser power density.

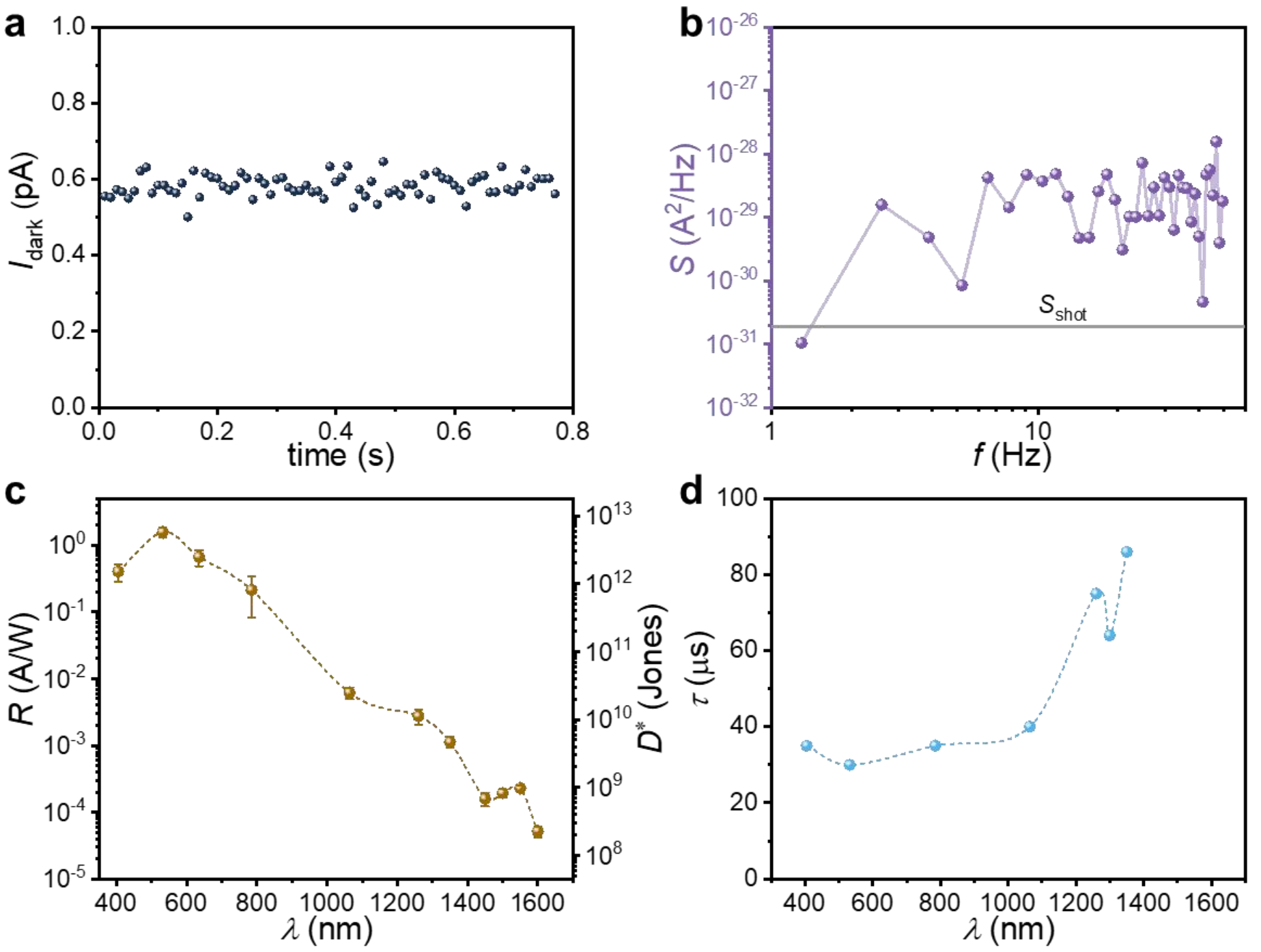

**Figure S9** **a** Dark current of Device B when $V_g = -50$ V, and the average value is 0.583 pA **b** Calculated noise spectral density based on dark current, and $S_{shot}$ denotes the shot noise component which is much smaller than the total noise. **c** $R$ and $D^*$, and **d** $\tau$ of the device versus wavelength from visible to NIR bands (from 405 to 1600 nm).

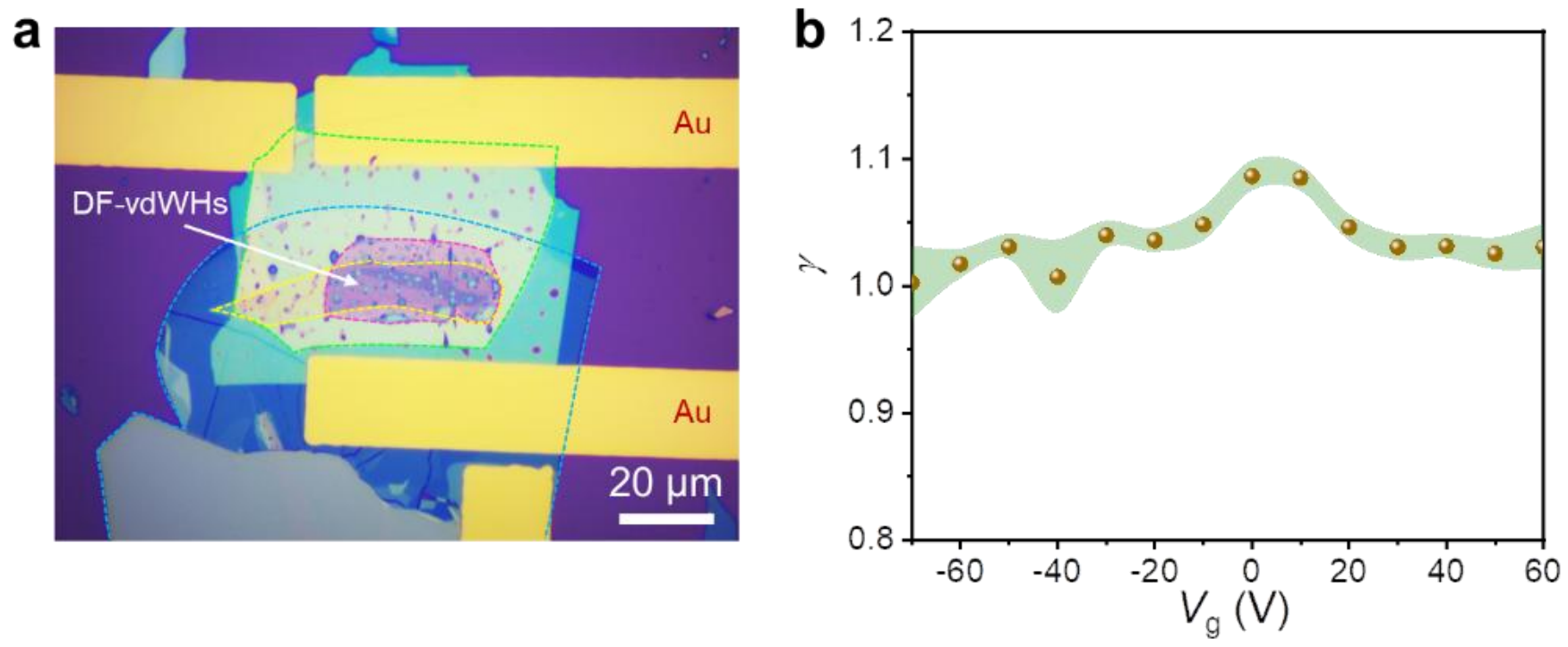

**Figure S10** The linearity of Device C at different gate voltage. **a** The optical image of the device. **b** The fitted linearities (brown spheres) at different gate voltages based on the data of Figure 4a, the green region indicates the deviation.

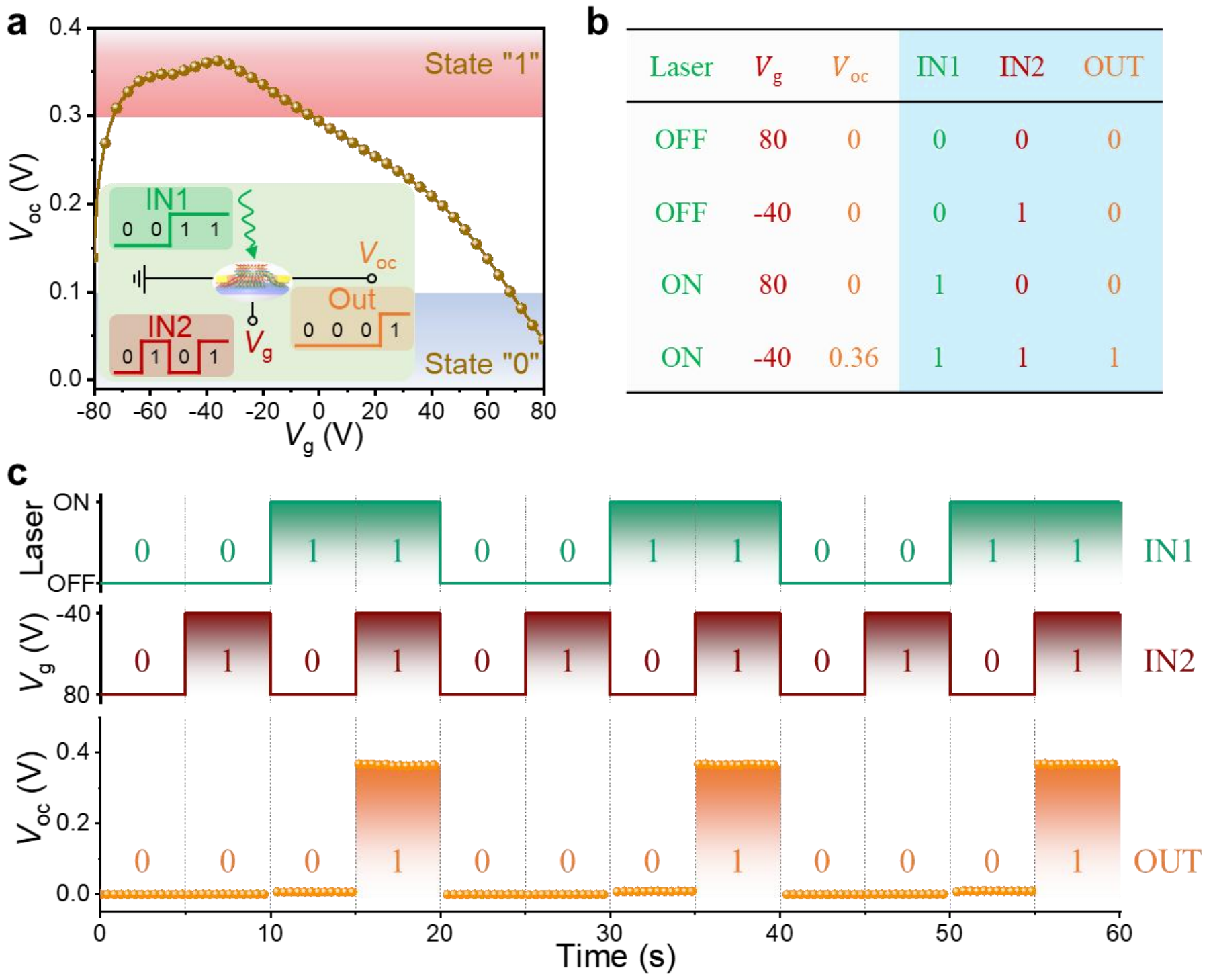

**Figure S11** Logic operation for optoelectronic AND gate with Device B. **a** $V_{oc}$ as a function of $V_g$ under 532 nm illumination when $P_{in}$ = 87.5 mW/cm$^2$. The states of $V_{oc}$ > 0.3 V and $V_{oc}$ < 0.1 V are defined as state "1" and state "0" for output, respectively. Inset: The schematic of the logic gate circuit. **b** Real table (left panel) and truth table (right panel) of AND Gate. **c** Operation result of the AND gate.

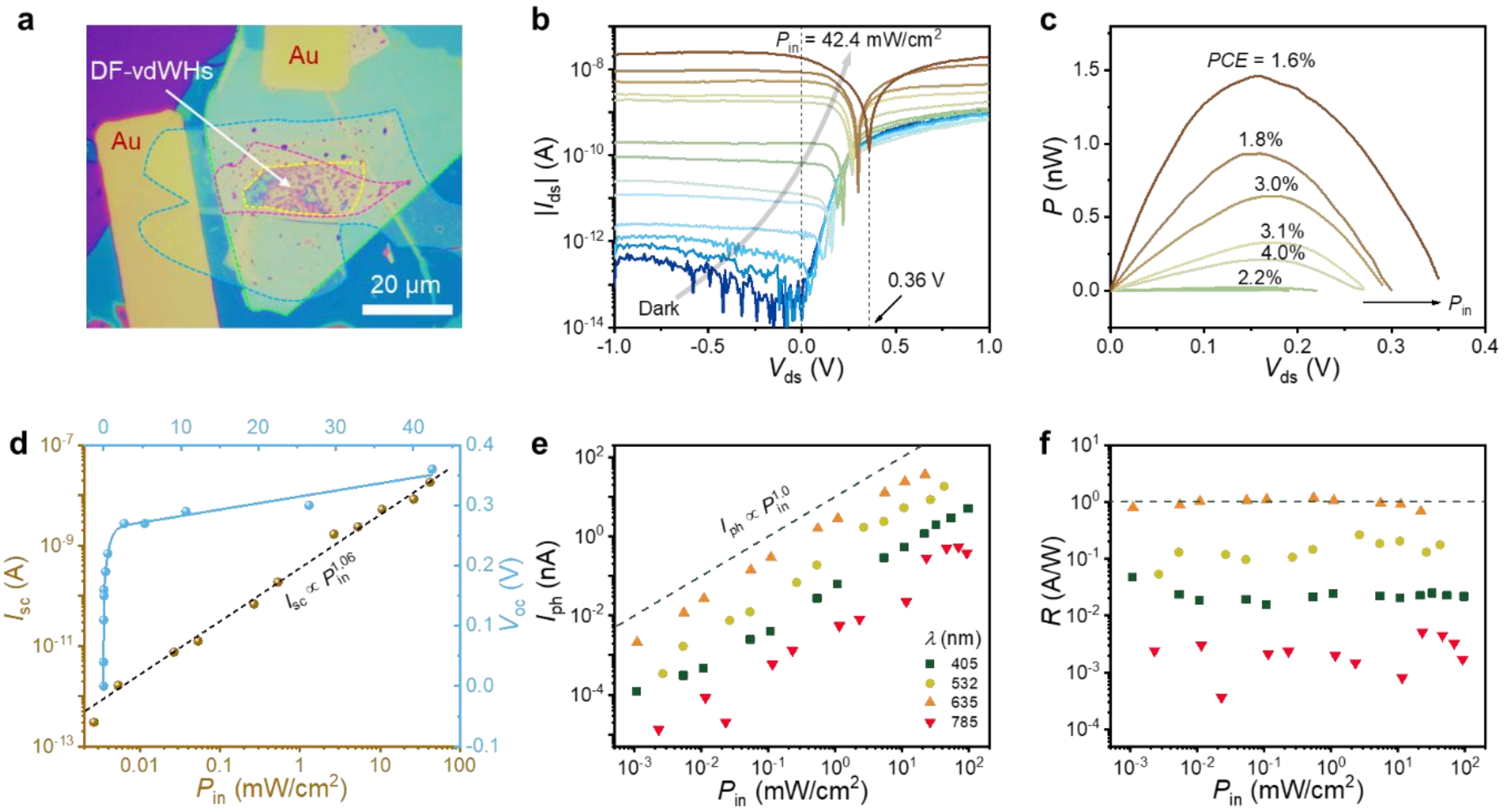

**Figure S12** Photodetection performance of the $MoS_2/WSe_2/MoS_2/WSe_2$ dual-floating vdWHs photodiode when $V_g = -60$ V. **a** Optical micrograph of the device, and layers from top to bottom are, $MoS_2$ (yellow), $WSe_2$ (blue), $MoS_2$ (green), $WSe_2$ (pink), and hBN, respectively, and the area of the dual-floating vdWHs is about 242 μm². **b** Output curves under 532 nm illumination with different $P_{in}$. **c**, **d** Extracted electrical power, $I_{sc}$ and $V_{oc}$ from output curves. **e**, **f** $I_{ph}$ and $R$ versus $P_{in}$ for the wavelength at 405, 532, 635, and 785 nm.

**Table S1** The thicknesses of each layer and the effective area of three $MoS_2/MoTe_2/MoS_2/MoTe_2$ dual-floating vdWHs photodiodes.

| Device | Thickness (nm) | | | | $A$ ($\mu m^2$) |
|---|---|---|---|---|---|
| | $MoS_2$ (4) | $MoTe_2$ (3) | $MoS_2$ (2) | $MoTe_2$ (1) | |
| Device A | 3.5 | 3.6 | 6.5 | 48 | 42.2 |
| Device B | 4 | 11 | 11 | 28 | 123.5 |
| Device C | 6 | 9 | 11 | 26 | 360 |

**Table S2** Comparison of figures of merit for $MoTe_2/MoS_2$ and $WSe_2/MoS_2$ photodetectors

| Materials | Linearity | Layer (nm) | $V_{oc}$ (V) | $\lambda$ (nm) | $R$ (mA/W) | $D^*$ (Jones) | Response time | Ref. |
|---|---|---|---|---|---|---|---|---|
| $MoTe_2/MoS_2$ Dual-floating | Yes | 4/11/11/28 | 0.404 | 405–1600 | 1570 (532 nm) | $4.28\times10^{11}$ | 30 μs | This work |
| $WSe_2/MoS_2$ Dual-floating | Yes | - | 0.36 | 405-785 | 1000 (635 nm) | - | 40 μs | This work |
| $WSe_2/MoS_2$ Single-floating | No | 41/12/8 | 0.34 | 405 | 715 | $1.59\times10^{13}$ | 45 μs | [3] |
| $MoTe_2/MoS_2$ | Yes | 7/6 | 0.45 | 405–1310 | 620 (532 nm) | $2\times10^{11}$ | 10 μs | [4] |
| $MoTe_2/MoS_2$ | - | 1.5/3.8 | 0.17 | 473 | 64 | $1.6\times10^{10}$ | 385 ms | [5] |
| $MoTe_2/MoS_2$ | - | 2.2/3 | 0.3 | 470–800 | 322 (470 nm) | - | 1 ms | [6] |
| $MoTe_2/MoS_2$ | No | 3.3/7 | 0.51 | 550–1550 | 46 (637 nm) | $1.06\times10^{8}$ | 60 μs | [7] |
| $MoTe_2/MoS_2$ | No | 6.4/13.7 | 0.15 | 473 | - | - | 1.1 ms | [8] |
| $MoTe_2/MoS_2$ | No | 8.5/6.5 | - | 450-980 | 146 (520 nm) | $2\times10^{11}$ | 172 μs | [9] |
| $WSe_2/MoS_2$ | Yes | 30/10 | 0.4 | 405–980 | 120 | - | 1 ms | [10] |
| $WSe_2/MoS_2$ | No | 11/1L | 0.26 | 532 | 0.27 | $2\times10^{11}$ | 4.1 μs | [11] |
| $WSe_2/MoS_2$ | No | 30/26.8 | 0.19 | 405 | 5390 | - | >16 μs | [12] |
| $WSe_2/MoS_2$ | Yes | 37/17 | 0.3 | 450-980 | 550 (532 nm) | $10^{11}$ | ~20 μs | [13] |

## References

[1] H. Li, Q. Zhang, C. C. R. Yap, B. K. Tay, T. H. T. Edwin, A. Olivier, D. Baillargeat, *Adv. Funct. Mater.* **2012**, 22, 1385.

[2] C. Ruppert, O. B. Aslan, T. F. Heinz, *Nano Lett.* **2014**, 14, 6231.

[3] Y. Jiang, R. Wang, X. Li, Z. Ma, L. Li, J. Su, Y. Yan, X. Song, C. Xia, *ACS Nano* **2021**, 15, 14295.

[4] J. Ahn, J.-H. Kang, J. Kyhm, H. T. Choi, M. Kim, D.-H. Ahn, D.-Y. Kim, I.-H. Ahn, J. B. Park, S. Park, Y. Yi, J. D. Song, M.-C. Park, S. Im, D. K. Hwang, *ACS Appl. Mater. Interfaces* **2020**, 12, 10858.

[5] F. Wang, L. Yin, Z. X. Wang, K. Xu, F. M. Wang, T. A. Shifa, Y. Huang, C. Jiang, J. He, *Adv. Funct. Mater.* **2016**, 26, 5499.

[6] A. Pezeshki, S. Hossein, H. Shokouh, T. Nazari, K. Oh, S. Im, *Adv. Mater.* **2016**, 28, 3216.

[7] Y. Chen, X. Wang, G. Wu, Z. Wang, H. Fang, T. Lin, S. Sun, H. Shen, W. Hu, J. Wang, *Small* **2018**, 14, 1703293.

[8] L. Yin, F. Wang, R. Cheng, Z. Wang, J. Chu, Y. Wen, J. He, *Adv. Funct. Mater.* **2019**, 29, 1804897.

[9] S. Li, Z. He, Y. Ke, J. Guo, T. Cheng, T. Gong, Y. Lin, Z. Liu, W. Huang, X. Zhang, *Appl. Phys. Express* **2019**, 13, 015007.

[10] H. S. Lee, J. Ahn, W. Shim, S. Im, D. K. Hwang, *Appl. Phys. Lett.* **2018**, 113, 163102.

[11] B. Liu, X. Zhang, J. Du, J. Xiao, H. Yu, M. Hong, L. Gao, Y. Ou, Z. Kang, Q. Liao, Z. Zhang, Y. Zhang, *InfoMat.* **2022**, 4, e12282.

[12] J.-B. Lee, Y. R. Lim, A. K. Katiyar, W. Song, J. Lim, S. Bae, T.-W. Kim, S.-K. Lee, J.-H. Ahn, *Advanced Materials* **2019**, 31, 1904194.

[13] J. Ahn, J.-H. Kang, M.-C. Park, D. K. Hwang, *Opt. Lett.* **2020**, 45, 4531.